\documentclass[journal=nalefd,manuscript=letter,layout=twocolumn, citeautoscript]{achemso}

\setkeys{acs}{etalmode=truncate}
\setkeys{acs}{maxauthors=10}
\setkeys{acs}{articletitle=true}

\usepackage{amssymb}

\usepackage{epsfig}
\usepackage{graphicx}
\usepackage{amsmath}
\usepackage{bm}
\usepackage{color}
\usepackage[normalem]{ulem}

\title{Giant Fine Structure Splitting of the Bright Exciton in a Bulk MAPbBr$_3$ Single Crystal}

\author{Micha{\l} Baranowski}
\affiliation{Laboratoire National des Champs Magn\'etiques Intenses, UPR 3228, CNRS-UGA-UPS-INSA, Grenoble and Toulouse, France}
\alsoaffiliation{Department of Experimental Physics, Faculty of Fundamental Problems of Technology, Wroclaw University of Science and
Technology, Wroclaw, Poland}

\author{Krzysztof Galkowski}
\affiliation{Laboratoire National des Champs Magn\'etiques Intenses, UPR 3228, CNRS-UGA-UPS-INSA, Grenoble and Toulouse, France}
\alsoaffiliation{Institute of Physics, Faculty of Physics, Astronomy and Informatics, Nicolaus Copernicus University, 5th Grudziadzka St., 87-100 Torun, Poland}

\author{Alessandro Surrente}
\affiliation{Laboratoire National des Champs Magn\'etiques Intenses, UPR 3228, CNRS-UGA-UPS-INSA, Grenoble and Toulouse, France}

\author{Joanna Urban}
\affiliation{Laboratoire National des Champs Magn\'etiques Intenses, UPR 3228, CNRS-UGA-UPS-INSA, Grenoble and Toulouse, France}

\author{{\L}ukasz K{\l}opotowski}
\affiliation{Institute of Physics, Polish Academy of Sciences, al. Lotnikow 32/46, 02-668 Warsaw, Poland}

\author{Sebastian. Ma{\'c}kowski}
\affiliation{Institute of Physics, Faculty of Physics, Astronomy and Informatics, Nicolaus Copernicus University, 5th Grudziadzka St., 87-100 Torun, Poland}

\author{Duncan Kennedy Maude}
\affiliation{Laboratoire National des Champs Magn\'etiques Intenses, UPR 3228, CNRS-UGA-UPS-INSA, Grenoble and Toulouse, France}

\author{Rim Ben Aich}
\affiliation{Laboratoire de Physique des Mat{\'e}riaux : Structure et Propri{\'e}t{\'e}s, Facult{\'e} des Sciences de Bizerte, Universit{\'e} de Carthage, 7021 Zarzouna, Bizerte, Tunisia}

\author{Kais Boujdaria}
\affiliation{Laboratoire de Physique des Mat{\'e}riaux : Structure et Propri{\'e}t{\'e}s, Facult{\'e} des Sciences de Bizerte, Universit{\'e} de Carthage, 7021 Zarzouna, Bizerte, Tunisia}

\author{Maria Chamarro}
\affiliation{Sorbonne Universit{\'e}, CNRS-UMR 7588, Institut des NanoSciences de Paris, INSP,
4 place Jussieu, F-75005, Paris, France}

\author{Christophe Testelin}
\affiliation{Sorbonne Universit{\'e}, CNRS-UMR 7588, Institut des NanoSciences de Paris, INSP,
4 place Jussieu, F-75005, Paris, France}

\author{Pabitra Nayak}
\affiliation{University of
Oxford, Clarendon Laboratory, Parks Road, Oxford, OX1 3PU, United
Kingdom}

\author{Markus Dollmann}
\affiliation{University of
Oxford, Clarendon Laboratory, Parks Road, Oxford, OX1 3PU, United
Kingdom}

\author{Henry James Snaith}
\affiliation{University of
Oxford, Clarendon Laboratory, Parks Road, Oxford, OX1 3PU, United
Kingdom}

\author{Robin Nicholas}
\email{robin.nicholas@physics.ox.ac.uk}
\phone{01865 (2)72250}
\affiliation{University of
Oxford, Clarendon Laboratory, Parks Road, Oxford, OX1 3PU, United
Kingdom}

\author{Paulina Plochocka}\email{paulina.plochocka@lncmi.cnrs.fr}\phone{+33 05 62 17 28 62}
\affiliation{Laboratoire National des Champs Magn\'etiques Intenses, UPR 3228, CNRS-UGA-UPS-INSA, Grenoble and Toulouse, France}
\alsoaffiliation{Department of Experimental Physics, Faculty of Fundamental Problems of Technology, Wroclaw University of Science and
Technology, Wroclaw, Poland}

\begin{document}


\begin{abstract}


\textit{Exciton fine structure splitting in semiconductors reflects the underlying symmetry of the crystal and quantum confinement. Since the latter factor strongly enhances the exchange interaction, most work has focused on nanostructures. Here, we report on the first observation of the bright exciton fine structure splitting in a bulk semiconductor crystal, where the impact of quantum confinement can be specifically excluded, giving access to the intrinsic properties of the material. Detailed investigation of the exciton photoluminescence and reflection spectra of a bulk methylammonium lead tribromide single crystal reveals a zero magnetic field splitting as large as $\sim 200\mu$\,eV.
This result provides an important starting point for the discussion of the origin of the large bright exciton fine structure observed in perovskite nanocrystals.}


\end{abstract}

\maketitle

In an ideally pure semiconductor, the lowest energy electronic excitation is a bound electron-hole pair (exciton). The exchange interaction between electron and the hole spins lifts the degeneracy between dark singlet and bright multiplet
excitonic states producing a fine structure. The physics of the fine structure splitting (FSS) has been the subject of intense
investigation\cite{andreani1990exchange,blackwood1994exchange,salmassi1989exchange,gammon1996fine,bayer1999electron,bayer2002fine,puls1999magneto,besombes2000exciton,nirmal1995observation,chen1988exchange,bester2003pseudopotential,mrowinski2015magnetic,mrowinski2016excitonic},
since it can be used for quantum logic\cite{li2003all}, and wave function control \cite{bonadeo1998coherent}. Moreover, the
ability to control the FSS is essential to fabricate the efficient entangled photon sources \cite{stevenson2006semiconductor,
bennett2010electric, benson2000regulated, muller2014demand,dousse2010ultrabright} required for quantum teleportation
\cite{bouwmeester1997experimental}, quantum communication \cite{briegel1998quantum}, and quantum cryptography \cite{gisin2002quantum}.

The pattern of excitonic states produced by the exchange interaction strongly depends on the symmetry of the system. In structures with sufficiently low symmetry, the degeneracy of the bright states should be completely lifted. However, in general, it
is expected that the bright exciton FSS is much smaller than the bright--dark state splitting \cite{bayer2002fine}, which is typically a few tens of
$\mu$eV in common inorganic semiconductors\cite{fu1999excitonic}. Since the symmetry breaking is generally also small (or even absent), the bright exciton
FSS has never been observed in any bulk semiconductor. In contrast, the bright exciton FSS is commonly observed in semiconductor
nanostructures\cite{andreani1990exchange,blackwood1994exchange,salmassi1989exchange,gammon1996fine,
bayer1999electron,bayer2002fine,puls1999magneto,besombes2000exciton} where quantum confinement greatly enhances the exchange interaction and the symmetry is strongly broken by the anisotropic quantum confinement.

Recently, different theoretical models predict that the low symmetry of lead-halide perovskites in the orthorombic phase should result in a significant bright exciton fine structure splitting. While there is no consensus about the origin of this phenomenon, which can result from the Rashba effect\cite{becker2018bright, isarov2017rashba, Sercel2019} or interplay of  exchange interaction and crystal field\cite{fu2017neutral, ramade2018fine, aich2019bright, Tamarat2019}, all of those predictions make lead halide perovskites a promising starting point to search for the elusive bright exciton FSS in bulk semiconductors.


Here, we report that indeed the bright exciton FSS can be observed in a high quality bulk MAPbBr$_3$ single crystal in which all the extrinsic sources of symmetry breaking and quantum confinement enhancement can be specifically excluded. Therefore our work constitutes an important base for further discussions of the intrinsic origin of exciton FSS in perovskite crystals.
By means of detailed magneto-optical spectroscopy, we reveal the FSS of the
exciton 1s transition with a splitting as large as 200$\mu$eV. We show that the observed value can be reasonably explained by the recent model proposed in refs.\cite{aich2019bright, ramade2018fine} with parameters extracted from our experimental studies.


\begin{figure}[h!]
\centering
\includegraphics[width=0.8\linewidth]{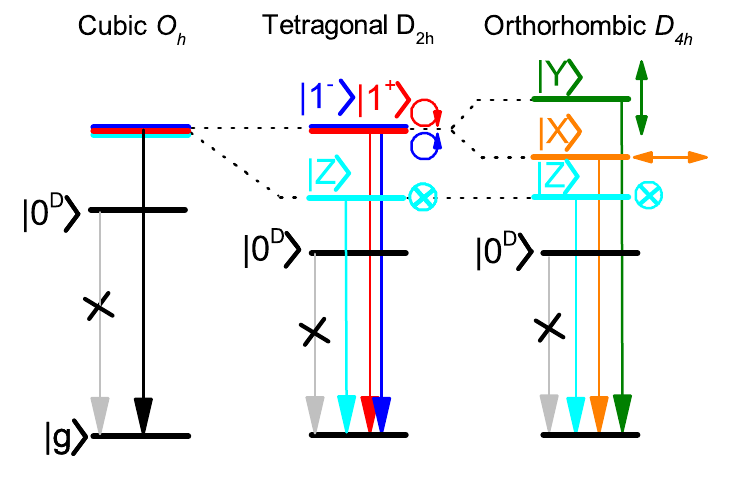}
\caption{ Schematic showing the exciton fine structure for the different crystal structures of perovskite and expected circular or linear polarization of the optical transitions. The presented scheme takes into account the exciton symmetry in the cubic ($O_{h}$), tetragonal ($D_{2h}$) and orthorombic ($D_{4h}$) cases, the crystal field and the exchange interaction\cite{fu2017neutral, ramade2018fine}. Including Rashba splittings could lead to further symmetry reduction and to the bright states being lower in energy than their dark counterpart\cite{becker2018bright}.}
\label{fig1}
\end{figure}

In lead-halide perovskites, the conduction and valence bands are built mainly from cationic Pb orbitals. Crucially, the valence
band is built from s-like orbitals, while the conduction band is built from p-like orbitals\cite{even2013importance, even2014dft,
brivio2014relativistic, endres2016valence}. Strong spin-orbit coupling in the conduction band\cite{even2013importance} splits the
electron states with total angular momentum $J=3/2$ (upper band) and $J=1/2$ (lower band). Therefore, the band-edge exciton can
be in one of four degenerate states: $\big |0_D\big >$ with zero total angular momentum, $J^{exc}=0$, and three states $\big|Z\big> $ , $\big |\pm 1\big >$ with $J^{exc}=1$ and $z$ components of angular momentum $J^{exc}_z=0$ or $J^{exc}_z=\pm 1$
respectively.

\begin{figure*}[th]
\centering
\includegraphics{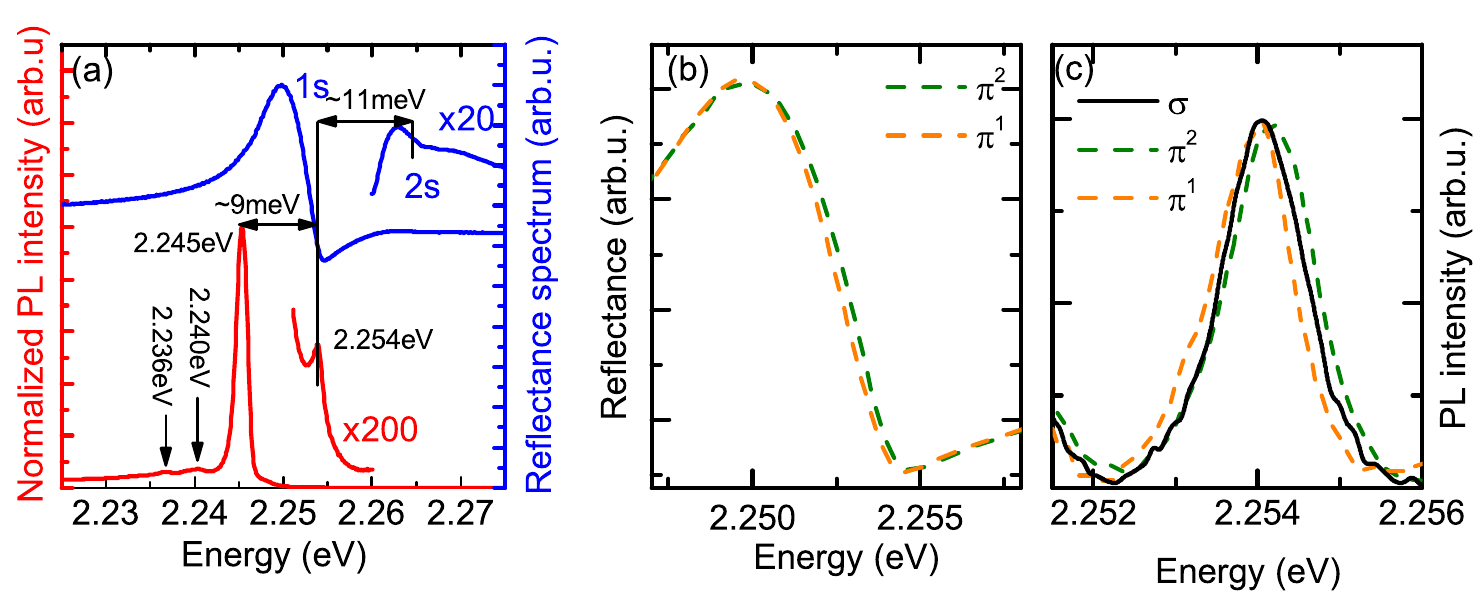}
\caption{PL (red line) and reflectance (blue line) spectra measured at $T = 1.5$\,K. The 1s excitonic transition is observed in
PL and both the 1s and 2s states can be identified in reflectance. The strong PL emission 9\,meV below the 1s transition (with no
corresponding feature in reflectance) is assigned to bound exciton emission. (b) Reflectance spectrum measured in a linear
polarization basis for two orthogonal polarizations. (c) Two linear orthogonally polarized components of PL spectra (dashed lines)
together with PL spectra measured in circular polarization basis. For clarity the PL background has been subtracted.}
\label{fig:PLandR}
\end{figure*}

In a cubic crystal, the exchange interaction splits the dark, $J^{exc}=0$, exciton state from the triply degenerate bright,
$J^{exc}=1$, states. Depending on temperature, the lead-halide perovskite crystal can be in a cubic, or tetragonal or orthorhombic
phase \cite{poglitsch1987dynamic, chen2018elucidating}. In the latter two phases, crystal field together with exchange interaction lifts the degeneracy of bright states\cite{fu2017neutral,odenthal2017spin,yu2016effective, ramade2018fine, nestoklon2018optical}. In the tetragonal phase, the bright exciton levels split into two
degenerate $\big |\pm 1\big >$ states, which couple to circularly (electric vector in $xy$ plane) polarized light and a $\big
|Z\big >$ state, which couples to linearly (electric vector along $z$ direction) polarized light. The further symmetry reduction in the orthorhombic (low temperature) phase lifts the degeneracy of the $\big |\pm 1\big >$ states resulting in
symmetric and antisymmetric states $\big|X\big >=\big(\big |1^+\big >+\big |1^-\big >\big)/\sqrt 2$, $\big|Y\big
> =\big(\big |1^+\big >-\big |1^-\big >\big)/\sqrt 2$, which couple to the two other linearly orthogonal polarizations of light, respectively. The expected fine structures of excitons in different crystallographic phases are summarized in Figure\,\ref{fig1}.

Very recently, bright exciton FSS  (in the range of
few hundred $\mu$eV up to 1-2\,meV) has been reported for lead halide perovskite nanocrystals\cite{yin2017bright, isarov2017rashba, fu2017neutral, becker2018bright, ramade2018fine, nestoklon2018optical, pfingsten2018phonon, raino2016single, ramade2018exciton}.
However it is not clear if the observed splitting is related only to the crystal field and exchange interaction\cite{yin2017bright,fu2017neutral, ramade2018fine, nestoklon2018optical} or if it is enhanced or driven by the Rashba effect\cite{becker2018bright,isarov2017rashba}.
Moreover, in nanocrystals one has to worry about possible contributions of anisotropic quantum confinement to the FSS \cite{nestoklon2018optical,ramade2018fine,bester2003pseudopotential,zielinski2013excitonic, seguin2005size, inprep2} or a surface enhancement of the Rashba effect\cite{niesner2016giant,mosconi2017rashba}.
The significant variation in the size of bright exciton FSS observed in the nanocrystals \cite{becker2018bright, yin2017bright} indicates that the fine structure splittings in perovskites can be strongly affected by extrinsic contributions. This hinders the determination of the intrinsic properties of perovskite crystals, which are required to provide a starting point for the discussion of the origin of the large bright exciton fine structure splitting. Moreover, the random orientation of the nanocrystals, together with the presence of high temperature phases even at cryogenic temperatures,\cite{swarnkar2016quantum} further complicates the interpretation of the exciton FSS.


The few mm by few mm single MAPbBr$_3$ crystal were prepared as reported for MAPbBr$_3$ in ref.\cite{nayak2016mechanism}. Briefly, MABr (from Dyesol) and PbBr$_2$ (Sigma-Aldrich) salts were added to dimethyl formamide (DMF) to prepare a 1M solution. Formic acid ($\sim$40 $\mu$l/ml) was added to the above solution followed by a filtration with a 0.45 micron filter. The solution incubated at ~55$^\circ$C produced multiple seed crystals of ($\sim$500$\mu$m). The seed crystals (which are  single crystalline as we have used them for XRD characterization  \cite{nayak2016mechanism}) were collected and used as seed for the growth of mm size crystals where the DMF solution has less formic acid. Each seed produced only one large crystal and the number of large crystals were controlled by the number of seeds added. Namely the filtered solution was poured into a glass vial and a cleaned glass slide was placed at the bottom of the vial. A few small seeds ($\sim$500$\mu$m) of MAPbBr$_3$ crystal were carefully placed on the glass slide. The glass vial was then incubated at $\sim^{\circ}$55C to produce the large  MAPbBr$_3$ crystals of size $2\times 2\times 1$\,mm$^3$. The crystals were collected when the solution is still hot ($\sim$55$^{\circ}$C) and preserved in chlorobenzene. Before the measurements, the crystals were blow dried with a N$_2$ gun. The measurements were performed on the largest facet of the crystal.

The  magnetic field measurements were performed in a cryofree split coil (up to 7\,T) equipped with a variable temperature
insert ($1.5-300$\,K). The optical access to the sample was through quartz B optical windows. The white light used for the
reflection studies was provided by a tungsten lamp. For the PL measurements the sample was excited using a 405\,nm continuous
wave (CW) laser with a power of 100$\mu$W. The PL and reflection measurement were performed in the Faraday geometry where the
excitation beam or white light was directed onto the sample through a non-polarizing beam splitter and focused on the sample by a
20\,cm focal length lens. The signal was collected by the same lens and dispersed by a 0.5\,m long spectrometer with a 1800
groove/mm grating. The dispersed light was detected using a nitrogen cooled CCD camera. The polarization of the signal was
analyzed using a broad band half wave or quarter wave plate and linear polarizer.


All our measurements have been performed in the low temperature orthorhombic phase(which was confirmed by temperature dependence of PL presented in SI). The macroPL and reflectance
spectrum of the MAPbBr$_3$ crystal taken at $T=1.5$\,K are presented in Figure \ref{fig:PLandR}(a). The PL spectrum (red curve) is dominated by
a sharp feature at 2.245\,eV with a 1.5\,meV full width at half maximum (FWHM).  In reflectance, in addition to the strong dispersive resonance observed
for the 1s free exciton transition, a much weaker feature corresponding to the 2s free exciton state is observed on the high
energy side (see expanded $\times 20$ curve). For the dispersive reflectance line shape the resonance position,
indicated by the vertical black dash/dotted lines, corresponds to the maximum slope.

The 1s and 2s states are separated by $\simeq 11$\,meV, consistent with an exciton binding energy $R_y^* \simeq 14.5$\,meV
(assuming a hydrogenic model). This value of exciton binding energy is in good agreement with those reported previously \cite{tilchin2016hydrogen}.
The dominant PL peak is red-shifted by $\simeq 9$\,meV
with respect to the resonance observed in the reflectance spectrum. Furthermore, on the high energy side of the PL spectrum there
is a relatively weak peak (see expanded $\times 200$ curve) at the same energy as the 1s excitonic transition observed in
reflectance. Therefore, we attribute the strong PL emission to bound states\cite{chen2018elucidating} (see also extended dissuasion in SI), and the small PL feature on the high energy side to the 1s free exciton emission.

Moreover two weak low energy peaks separated by about 5meV and 9 meV from the strongest PL peak can be observed. Such peaks have already been observed in CsPbBr$_3$ perovskite nanocrystals and attributed to TO phonon replicas \cite{fu2017neutral}. Interestingly the separation between the first phonon replica and the 1s excitonic peak is 14meV which is also very close to the effective energy of the LO phonon reported for MAPbBr$_3$ crystals \cite{wright2016electron}. Therefore the lowest energy bands can be attributed to phonon replicas of the bound and free exciton transitions.

Reflectance and PL spectra measured in a linear polarization basis are shown in Figure \ref{fig:PLandR}(b,c). The energy shift between the orthogonal linear polarizations $\pi^1$ and $\pi^2$, clearly visible in both PL and reflectance, is the smoking gun signature of the bright exciton FSS in a bulk crystal.


\begin{figure*}[t]
\centering
\includegraphics{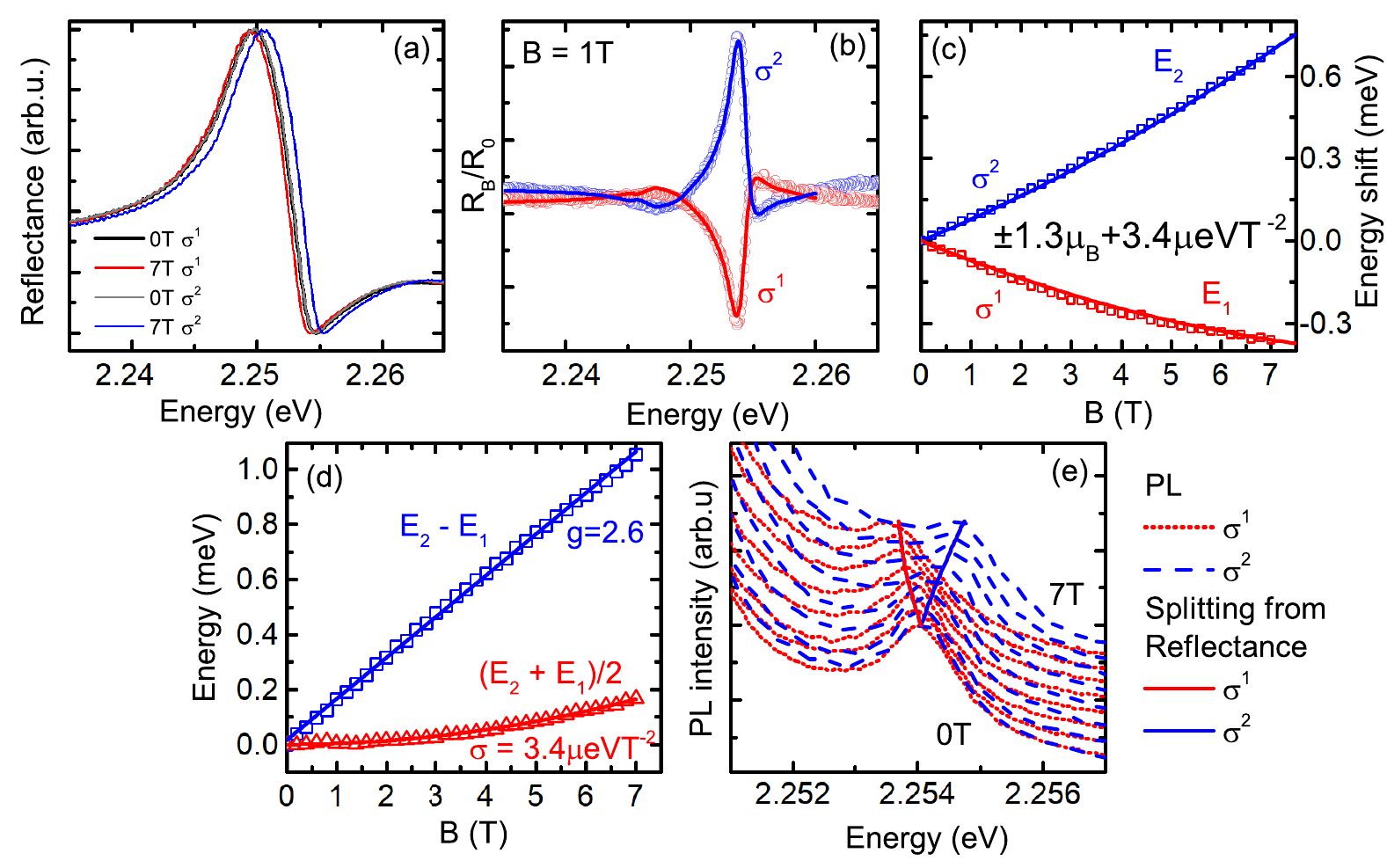}
\caption{(a) $\sigma^1$ and $\sigma^2$ components of the reflectance spectrum measured at zero (black and gray curve) and 7\,T
(red and blue). (b) Reflectance spectrum at 1\,T divided by the zero field spectra (open points) together with a fit using
equation \eqref{equ:magic_fit}. (c) Shift of the $\sigma^1$ and $\sigma^2$ components as a function of magnetic field. (d)
Determined Zeeman splitting (blue squares) and diamagnetic shift (red triangles) together with linear and quadratic fit, respectively. (e) Comparison of the reflectance shift with the PL spectra feature assigned to free exciton emission.}
\label{fig:RvsB}
\end{figure*}

Before discussing in detail the zero magnetic field FSS, we present magneto-optical data taken in the Faraday configuration, which unambiguously demonstrates the free excitonic nature of the high energy PL feature. Figure\,\ref{fig:RvsB}(a) shows reflectance spectra measured for the two opposite circular polarizations $\sigma^1$ and $\sigma^2$ at 0 and 7\,T
magnetic field. The expected shift 
of the $\sigma^1$ and $\sigma^2$ components in magnetic
field is clearly seen. To determine precisely the splitting of the $\sigma^1$ and $\sigma^2$ transition as a function of magnetic
field we divide them by the zero field spectrum (see Figure \ref{fig:RvsB}(b)).
The ratioed spectra have a characteristic sharp feature with an amplitude and width which increase with the size of the energy shift. As the shape of the reflectance spectra are almost unchanged in magnetic field, the shift can be measured by fitting to the ratioed spectra by varying the shift between the spectra in
\begin{equation}
\frac{R_B (E)}{R_0(E)} \propto \frac{R_0(E+\Delta E)}{R_0(E)},
\label{equ:magic_fit}
\end{equation}
where $R_B (E)$ and $R_0(E)$ are reflectance spectra in magnetic field and at zero field and $\Delta E$ is the shift of the
reflectance resonance in magnetic field. More details about fitting procedure can be found in the supplementary information. The fits,
shown as solid lines Figure \ref{fig:RvsB}(b), are in excellent agreement with the experimental data. This method allows us to precisely determine the shift of the exciton transition even at very low magnetic field and avoids having to model the reflection line shape.

The energy shift of the $\sigma^1$ and $\sigma^2$ transitions and the resulting Zeeman splitting and diamagnetic shift are
summarized in Figure \ref{fig:RvsB}(c,d). Fitting the splitting of the $\sigma^1$ and $\sigma^2$ features with $E_2 - E_1=g\mu_B B$ (see Figure \ref{fig:RvsB}(d)) gives $g = 2.6 \pm 0.1$, in agreement with the previously reported value\cite{odenthal2017spin}. 
The diamagnetic shift is extracted by fitting the average behavior in field ($\sigma B^2 = (E_2 + E_1)/2$), giving $\sigma=(3.4 \pm
0.1)\,\mu$eVT$^{-2}$ with a resulting exciton Bohr radius of $\simeq 3.8$\,nm and effective mass of $\mu=0.185\,m_0$. 
Finally, we compare the behavior of the exciton transitions in magneto-reflectance with the behavior of the corresponding 1s features in PL (broken lines Figure \ref{fig:RvsB}(e))). The 1s PL feature shifts in exactly the same way as in the reflectance data, plotted in solid lines.

\begin{figure}[t]
\centering
\includegraphics[width=0.95\linewidth]{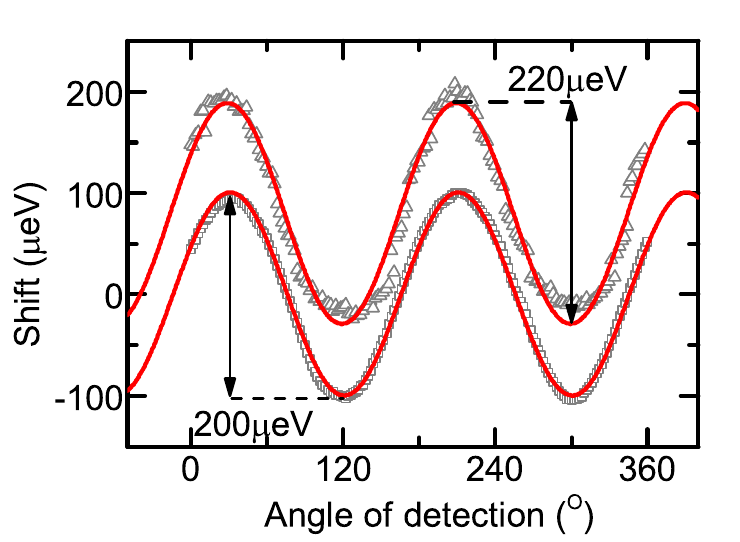}
\caption{(a) Shift of the reflectance resonance (open squares) and PL peak (open triangles) as a function of detection angle of
linear polarization. For clarity, the PL data is shifted vertically by 100$\mu$eV. The red lines are fits using a $\sin^2\theta$ function.}

\label{fig:linear}
\end{figure}

Two in-plane bright excitons should be linearly polarized and energetically separated (see also discussion in supplementary information).
To confirm that the zero magnetic field spectrum is composed of two linearly polarized components and to precisely determine the splitting
energy we have measured the energy shift versus the angle of the linear polarization in detection. The energy shift in
reflectance (open squares) and PL (open triangles) shows the expected oscillatory behavior with the angle of the polarization analyser (see Figure \ref{fig:linear}(a), where the PL data has been shifted vertically by 100$\mu$eV for clarity). The data is well described
using $\delta\sin^2(\phi+\phi_0)$ (red lines in Figure \ref{fig:linear}(a)), where $\delta$ is the bright exciton FSS, $\phi$ is the angle of linear polarization detection and $\phi_0$ is a phase factor. Similar values of the zero field splitting
are extracted from fitting the reflectance ($\delta \simeq 197\pm 7\mu$eV) and PL ($\delta \simeq 220\pm 40 \mu$eV).
A detailed discussion of uncertainty estimation can be found in the supplementary information.

\begin{figure}[h]
\centering
\includegraphics[width=0.95\linewidth]{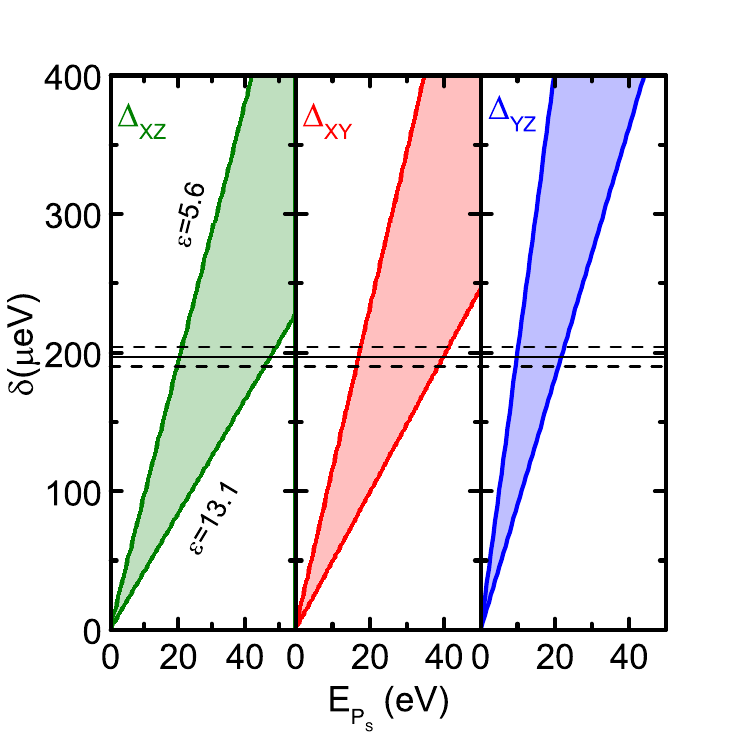}
\caption{Calculated bright exciton fine structure splittings as a function of the Kane energy $E_{P_S}$ for values of the dielectric constant $\epsilon_r$ from 5.6 to 13.1 (see text for details). The red, blue and green colors corresponds to $\Delta_{XY}$, $\Delta_{XZ}$ and $\Delta_{YZ}$ respectively. The experimentally observed splitting is indicated by the horizontal black line. The dashed lines indicate the estimated error for the reflectance measurements. Details of the calculations and numeric values of all used parameters can be found in supplementary information.}

\label{fig:calculation}
\end{figure}

Theoretically the quantitative evolution of exciton fine structure exchange splitting in different perovskite phases can be derived from symmetry considerations applied to the exciton picture \cite{fu2017neutral,yu2016effective, kataoka1993magneto, tanaka2005electronic}. Recently this approach has been used to provide a relation between the band structure parameters and the bright exciton fine structure splitting \cite{ramade2018fine}. The observed splittings between exciton states with dipole moment along X, Y and Z can be expressed as

\begin{equation}
   \Delta_{XY}=2J_b|\alpha^2-\beta^2|
   \label{equ:XY}
\end{equation}
\begin{equation}
   \Delta_{XZ}=2J_b|\beta^2-\gamma^2|
\end{equation}
\begin{equation}
   \Delta_{YZ}=2J_b|\alpha^2-\gamma^2|
\end{equation}
where the $J_b$ is an exchange integral and $\alpha$, $\beta$ and $\gamma$ are the weights of the Bloch wave functions of electrons in the orthorombic crystal structure. The exchange integral $J_b$ depends on the exciton Bohr radius and Kane energy ($E_{P_S}$), while $\alpha$, $\beta$ and $\gamma$ depend on the strength of the spin-orbit coupling and the crystal field (see detailed discussion in supplementary information).

From our magneto-optical measurements, we can extract the exciton Bohr radius (3.8 nm), the effective mass ($\mu=0.185$) and the dielectric constant ($\epsilon_r = 13.1$) which lies between the accepted high (5.6-6.7)\cite{ndione2016effects, valverde2015intrinsic, zhao2017low} and low frequency (25-38)\cite{zhao2017low, anusca2017dielectric} values. Such a discrepancy is typical for materials with smaller exciton radii\cite{miura2008physics}. Unfortunately, the Kane energy $E_{P_S}$ and $\alpha$ and $\beta$ parameters for MAPbBr$_3$ are not well established. Taking into account that the spin orbit-coupling and crystal field are expected to be similar for fully inorganic and hybrid perovskites\cite{ramade2018fine, yu2016effective} and that MAPbBr$_3$ and CsPbBr$_3$ have comparable band parameters\cite{galkowski2016determination, yang2017impact} we assume $\alpha=0.67$ and $\beta=0.57$ as in CsPbBr$_3$\cite{ramade2018fine}. The estimation of the Kane energy based on 4 and 40 band $\mathbf{k\cdot p}$ model gives $E_{P_S}=18.4$\,eV and 28.4\,eV (see supplementary information for details and ref. \cite{aich2019bright}). In the literature, one can also find values for the Kane energy in the range of $37.1-39.9$\,eV in the inorganic compounds CsPbX$_3$ (X = I, Br, Cl) \cite{becker2018bright, nestoklon2018optical}, although these would correspond to substantially lower effective mass values. The calculated value of the splitting between different excitonic states as a function of the Kane energy are presented in Figure\,\ref{fig:calculation} for values of the dielectric constant in the range $\epsilon_r = 5.6-13.1$.The predicted splitting of the bright excitonic states is in the range of a few hundreds of $\mu$eV in agreement with the experimental observation. Since all axes are equivalent in the room temperature cubic phase the orientation of the crystal axis cannot be determined after the transition to the low temperature orthorhombic phase. The accepted range of Kane energies ($\simeq$ 15-40\,eV) is however sufficient to correspond well with the calculated splitting for any of the three possible alignments.


\begin{figure}[h]
\centering
\includegraphics[width=0.95\linewidth]{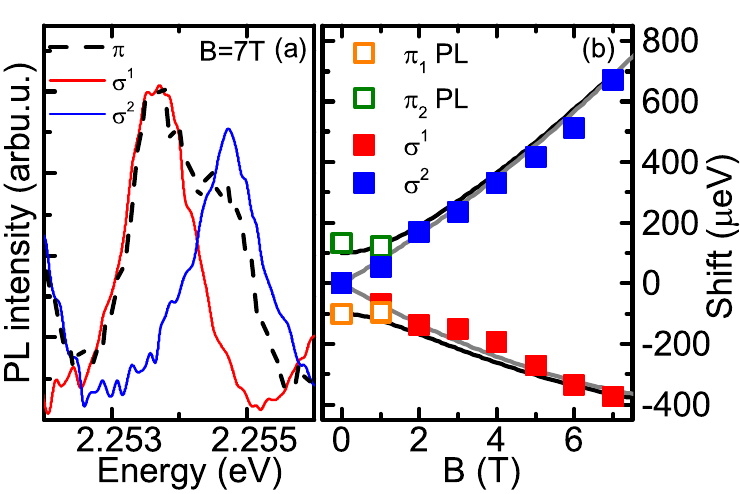}
\caption{(a) PL spectra measured at 7T in circular polarization basis (red and blue curves) and linear basis (dashed black curve). (b)
Dependence of PL peak position as a function of magnetic field measured in linear (open squares) and circular (full squares)
polarization basis. The black lines are calculated using the exchange Hamiltonian with parameters extracted from
the reflectance measurements $\delta=197\mu$eV, $g=2.6$ and $\sigma=3.4\mu$eVT$^{-1}$. The grey curve corresponding to the linear basis is calculated with $\delta=0$, \textit{i.e.}, neglecting the zero field splitting.}
\label{fig:magnetic2}
\end{figure}

We have also investigated the behaviour of the FSS in magnetic field. Figure \ref{fig:magnetic2}(a) shows PL spectra measured at $B=7$\,T in circular and linear polarization basis. In the circular basis splitting between states corresponding to $\sigma^1$ and $\sigma^2$ components of the emission is clearly visible, while both components are observed simultaneously in the linear basis. The detailed evolution of the PL peaks in the circular and linear polarization basis is shown in Figure \ref{fig:magnetic2}(b). The fine structure splitting of the 1s bright exciton in the orthorhombic phase, in the presence of magnetic field along the direction perpendicular to the dipole moment orientation, can be described in the ($\big |1^+\big>$, $\big |1^-\big>$) basis by the following Hamiltonian (see supplementary information and references \cite{fu2017neutral, yu2016effective}):

\begin{equation}
H=\frac{1}{2}\begin{pmatrix}
g \mu_B B+2\sigma B^2 & \delta\\
 \delta & -g\mu_B B+2\sigma B^2  \end{pmatrix}
\end{equation}
The corresponding eigenvalues and eigenstates are:
\begin{equation}
E_{Y/X}=\pm\frac{1}{2}\sqrt{(g\mu_BB)^2+{\delta}^2}+\sigma B^2 
\end{equation}

\begin{equation}
\big| X/Y\big>=A\Bigg[\big|+1\big>+\Bigg(- \frac{g\mu_BB}{\delta} \mp\sqrt{\frac{(g\mu_B B)^2}{\delta^2}+1}\,\Bigg) \big|-1\big >\Bigg]
\end{equation}
where $A$ is a field dependent normalizing constant. At zero magnetic field the energy difference between the states is equal to
$\delta$ and the states couple to linearly polarized light (superposition of states $\big|\pm 1\big>$). Increasing magnetic field induces a continuous change from linear polarization to circular polarization of the excitonic transitions in agreement with experiment (see (Figure \ref{fig:PLandR}(c)) and Figure \ref{fig:magnetic2}(a)). Finally, we compare
the dependence of the PL peak position as a function of the magnetic field  (see Fig.\ \ref{fig:magnetic2}(b)) with the predictions of
the exchange Hamiltonian (black lines) using the parameters extracted from reflectance studies, \emph{i.e.}, $\delta=200\mu$eV,
$g=2.6$ and $\sigma=3.4\mu$eVT$^{-1}$. The exchange Hamiltonian provides a full quantitative description of the data. At low magnetic field, the bright FSS is determined by the zero field splitting $\delta$, and with increasing magnetic field the transition energy is increasingly dominated by the Zeeman splitting.

It is worth noting that now one can find two interpretations of the splitting of the exciton states in perovskite nanocrystals. One is solely based on crystal symmetry, exchange interaction and crystal field \cite{fu2017neutral, ramade2018fine, aich2019bright, Tamarat2019} while the other invokes a Rashba effect to explian FSS in fully inorganic perovskite nanocrystals \cite{becker2018bright, Sercel2019, isarov2017rashba}. The Rashba effect\cite{becker2018bright} has often been invoked to explain several of the unique properties of the lead halide perovskites \cite{kepenekian2017rashba, zheng2015rashba, etienne2016dynamical, quarti2014interplay, mosconi2017rashba}. Despite the obvious beauty of this idea, it nevertheless remains highly controversial. For instance, the Rashba effect requires a lack of inversion symmetry or a breaking of bulk inversion symmetry, not present in centrosymmetric perovskite crystals. It was proposed that temporal symmetry breaking is provided by the cation motion, which leads to local polar fluctuations in mixed and fully inorganic perovskites \cite{yaffe2017local, mosconi2017rashba,etienne2016dynamical}. However, such vibrations of the cations should be absent at very low temperatures \cite{yaffe2017local, dar2016origin}, as used for all exciton fine structure studies to date. The absence of the Rashba effect in the orthorombic phase is also consistent with circular photogalvanic measurements performed on the analogue MAPbI$_3$ compound \cite{niesner2018structural}.

Another source of symmetry breaking might be attributed to surface effects \cite{niesner2016giant}, which can be especially important for nanocrystals providing a possible explanation for the widely varying (200-1000$\mu$eV) bright exciton FSS observed in the CsPbBr$_3$ nanocrystals. However, in our case, both the optical methods used here, namely reflectivity and PL, probe the bulk properties of the crystal. The absorption coefficient in MAPbBr$_3$ is of the order of $\alpha = 10^5$\,cm$^{-1}$ at a PL excitation wavelength of 405\,nm which corresponds to an optical absorption depth ($\alpha_z = 1$) of 100 nm. The exciton diffusion will further increase the volume sampled by the PL. In the presence of a very  large surface electric field the penetration depth sampled by reflectivity can fall to of the order of 20 nm ($\lambda$/(4$\pi n$) where $n$ is the real part of the refractive index)\cite{yang2015low, sabbah2002femtosecond}. This is still large on the scale of the lattice and exciton wavefunction ($\sim$4\,nm) and 2 times bigger than maximum surface height fluctuation amplitude (at most ±10nm over distances of order 100\,$\mu$m and much less on a wavelength scale, see SI). We have no evidence to suggest that such surface fields exist and it should be noted that the PL and reflectivity studies give identical results (within experimental error) in sign and magnitude when sampling different regions of the crystal. Therefore, both techniques measure the same property, namely the fine structure splitting of the bulk free exciton, which has not been significantly influenced by the surface.

Finally, according to models for the Rashba effect, the order of the bright and dark states should be reversed in fully inorganic perovskite nanocrystals\cite{becker2018bright, Sercel2019} which seems to be in contrast to some of the experimental reports on the PL dynamics in inorganic perovskite nanocrystals\cite{ramade2018fine, canneson2017negatively, chen2018elucidating, xu2018long} and recent direct observation of dark ground states in FAPbBr$_3$ nanocrystals\cite{Tamarat2019}. Given all above arguments and the good quantitative agreement of the theoretical predictions based only on crystal symmetry considerations’ this suggests that the Rashba effect plays little role in macroscopic MAPbBr$_3$ single crystals (and FAPbBr$_3$\cite{Tamarat2019}), at least for the lowest temperature phases.

It should be noted however that the excitonic band parameters, in particular the reduced effective mass, $\mu$, are significantly different from those deduced at higher energy from the magnetic field dependence of the nearly free electron Landau levels for thin films \cite{galkowski2016determination}. This discrepancy may be a hallmark of polaronic effects in perovskite crystals \cite{zhu2015charge, schlipf2018carrier, sendner2016optical, cinquanta2019ultrafast}. The effective mass extracted here comes from the low field limit while previous reports concentrate on free electron Landau levels observed in high field limit where the motion of carriers is decoupled from lattice vibration \cite{miura2008physics}. Therefore the previously extracted effective mass is lower than in the low field limits. However, this aspect is beyond the scope of this work and will be the subject of further studies.

In summary, we have observed a significant bright exciton fine structure splitting $\delta=197\,\mu$eV in bulk MAPbBr$_3$. This represents an important step in the understanding of the fine structure splitting in lead-halide perovskites and perovskite based nanostructures. Using a large single crystal, we are able to exclude the influence of surface, size and anisotropic confinement on the observed bright exciton FSS giving direct access to the intrinsic properties of the bulk MAPbBr$_3$. The observed splitting can be understood in the exciton picture combined with symmetry considerations. We show that, within this model, the observed splitting can be reasonably estimated based on band structure parameters derived from magneto-optical studies. Our work constitutes a firm base for further exciton fine structure studies in lead-halide perovskites. The presented resulst should provide valuable insight for future investigations of the different mechanisms that contribute to the exciton fine structure, such as confinement anisotropy or Rashba effect, in perovskite based nanostructures.

\begin{acknowledgement}
This work was partially supported by BLAPHENE and STRABOT projects, which received funding from the IDEX Toulouse, Emergence
program,  ``Programme des Investissements d'Avenir'' under the program ANR-11-IDEX-0002-02, reference ANR-10-LABX-0037-NEXT, and
by the PAN--CNRS collaboration within the PICS 2016-2018 agreement. We thank EPSRC for funding through grant EP/M05173/1. This work was supported by EPSRC also via its membership to the EMFL (Grant No.\ EP/N01085X/1). M.B. appreciates support from the Polish Ministry of Science
and Higher Education  within  the  Mobilnosc  Plus program (grant no.\ 1648/MOB/V/2017/0).
This work was partially supported by the French National Research Agency (ANR IPER-Nano2, ANR-18-CE30-0023-01)
\end{acknowledgement}

\section*{Supporting Information}
PL temperature dependence, AFM surface roughens measurements, uncertainty analysis, derivation of electron-hole exchange interaction.

\bibliography{FSS}

\providecommand{\latin}[1]{#1}
\makeatletter
\providecommand{\doi}
  {\begingroup\let\do\@makeother\dospecials
  \catcode`\{=1 \catcode`\}=2 \doi@aux}
\providecommand{\doi@aux}[1]{\endgroup\texttt{#1}}
\makeatother
\providecommand*\mcitethebibliography{\thebibliography}
\csname @ifundefined\endcsname{endmcitethebibliography}
  {\let\endmcitethebibliography\endthebibliography}{}
\begin{mcitethebibliography}{78}
\providecommand*\natexlab[1]{#1}
\providecommand*\mciteSetBstSublistMode[1]{}
\providecommand*\mciteSetBstMaxWidthForm[2]{}
\providecommand*\mciteBstWouldAddEndPuncttrue
  {\def\EndOfBibitem{\unskip.}}
\providecommand*\mciteBstWouldAddEndPunctfalse
  {\let\EndOfBibitem\relax}
\providecommand*\mciteSetBstMidEndSepPunct[3]{}
\providecommand*\mciteSetBstSublistLabelBeginEnd[3]{}
\providecommand*\EndOfBibitem{}
\mciteSetBstSublistMode{f}
\mciteSetBstMaxWidthForm{subitem}{(\alph{mcitesubitemcount})}
\mciteSetBstSublistLabelBeginEnd
  {\mcitemaxwidthsubitemform\space}
  {\relax}
  {\relax}

\bibitem[Andreani and Bassani(1990)Andreani, and Bassani]{andreani1990exchange}
Andreani,~L.~C.; Bassani,~F. Exchange interaction and polariton effects in
  quantum-well excitons. \emph{Phys. Rev. B} \textbf{1990}, \emph{41},
  7536\relax
\mciteBstWouldAddEndPuncttrue
\mciteSetBstMidEndSepPunct{\mcitedefaultmidpunct}
{\mcitedefaultendpunct}{\mcitedefaultseppunct}\relax
\EndOfBibitem
\bibitem[Blackwood \latin{et~al.}(1994)Blackwood, Snelling, Harley, Andrews,
  and Foxon]{blackwood1994exchange}
Blackwood,~E.; Snelling,~M.; Harley,~R.; Andrews,~S.; Foxon,~C. Exchange
  interaction of excitons in GaAs heterostructures. \emph{Phys. Rev. B}
  \textbf{1994}, \emph{50}, 14246\relax
\mciteBstWouldAddEndPuncttrue
\mciteSetBstMidEndSepPunct{\mcitedefaultmidpunct}
{\mcitedefaultendpunct}{\mcitedefaultseppunct}\relax
\EndOfBibitem
\bibitem[Salmassi and Bauer(1989)Salmassi, and Bauer]{salmassi1989exchange}
Salmassi,~B.~R.; Bauer,~G.~E. Exchange interaction in type-II quantum wells.
  \emph{Phys. Rev. B} \textbf{1989}, \emph{39}, 1970\relax
\mciteBstWouldAddEndPuncttrue
\mciteSetBstMidEndSepPunct{\mcitedefaultmidpunct}
{\mcitedefaultendpunct}{\mcitedefaultseppunct}\relax
\EndOfBibitem
\bibitem[Gammon \latin{et~al.}(1996)Gammon, Snow, Shanabrook, Katzer, and
  Park]{gammon1996fine}
Gammon,~D.; Snow,~E.; Shanabrook,~B.; Katzer,~D.; Park,~D. Fine structure
  splitting in the optical spectra of single GaAs quantum dots. \emph{Phys.
  Rev. Lett.} \textbf{1996}, \emph{76}, 3005\relax
\mciteBstWouldAddEndPuncttrue
\mciteSetBstMidEndSepPunct{\mcitedefaultmidpunct}
{\mcitedefaultendpunct}{\mcitedefaultseppunct}\relax
\EndOfBibitem
\bibitem[Bayer \latin{et~al.}(1999)Bayer, Kuther, Forchel, Gorbunov, Timofeev,
  Sch{\"a}fer, Reithmaier, Reinecke, and Walck]{bayer1999electron}
Bayer,~M.; Kuther,~A.; Forchel,~A.; Gorbunov,~A.; Timofeev,~V.;
  Sch{\"a}fer,~F.; Reithmaier,~J.; Reinecke,~T.; Walck,~S. Electron and hole g
  factors and exchange interaction from studies of the exciton fine structure
  in In 0.60 Ga 0.40 As quantum dots. \emph{Phys. Rev. Lett.} \textbf{1999},
  \emph{82}, 1748\relax
\mciteBstWouldAddEndPuncttrue
\mciteSetBstMidEndSepPunct{\mcitedefaultmidpunct}
{\mcitedefaultendpunct}{\mcitedefaultseppunct}\relax
\EndOfBibitem
\bibitem[Bayer \latin{et~al.}(2002)Bayer, Ortner, Stern, Kuther, Gorbunov,
  Forchel, Hawrylak, Fafard, Hinzer, Reinecke, \latin{et~al.}
  others]{bayer2002fine}
Bayer,~M.; Ortner,~G.; Stern,~O.; Kuther,~A.; Gorbunov,~A.; Forchel,~A.;
  Hawrylak,~P.; Fafard,~S.; Hinzer,~K.; Reinecke,~T. \latin{et~al.}  Fine
  structure of neutral and charged excitons in self-assembled In (Ga) As/(Al)
  GaAs quantum dots. \emph{Phys. Rev. B} \textbf{2002}, \emph{65}, 195315\relax
\mciteBstWouldAddEndPuncttrue
\mciteSetBstMidEndSepPunct{\mcitedefaultmidpunct}
{\mcitedefaultendpunct}{\mcitedefaultseppunct}\relax
\EndOfBibitem
\bibitem[Puls \latin{et~al.}(1999)Puls, Rabe, W{\"u}nsche, and
  Henneberger]{puls1999magneto}
Puls,~J.; Rabe,~M.; W{\"u}nsche,~H.-J.; Henneberger,~F. Magneto-optical study
  of the exciton fine structure in self-assembled CdSe quantum dots.
  \emph{Phys. Rev. B} \textbf{1999}, \emph{60}, R16303\relax
\mciteBstWouldAddEndPuncttrue
\mciteSetBstMidEndSepPunct{\mcitedefaultmidpunct}
{\mcitedefaultendpunct}{\mcitedefaultseppunct}\relax
\EndOfBibitem
\bibitem[Besombes \latin{et~al.}(2000)Besombes, Kheng, and
  Martrou]{besombes2000exciton}
Besombes,~L.; Kheng,~K.; Martrou,~D. Exciton and biexciton fine structure in
  single elongated islands grown on a vicinal surface. \emph{Phys. Rev. Lett.}
  \textbf{2000}, \emph{85}, 425\relax
\mciteBstWouldAddEndPuncttrue
\mciteSetBstMidEndSepPunct{\mcitedefaultmidpunct}
{\mcitedefaultendpunct}{\mcitedefaultseppunct}\relax
\EndOfBibitem
\bibitem[Nirmal \latin{et~al.}(1995)Nirmal, Norris, Kuno, Bawendi, Efros, and
  Rosen]{nirmal1995observation}
Nirmal,~M.; Norris,~D.~J.; Kuno,~M.; Bawendi,~M.~G.; Efros,~A.~L.; Rosen,~M.
  Observation of the" dark exciton" in CdSe quantum dots. \emph{Phys. Rev.
  Lett.} \textbf{1995}, \emph{75}, 3728\relax
\mciteBstWouldAddEndPuncttrue
\mciteSetBstMidEndSepPunct{\mcitedefaultmidpunct}
{\mcitedefaultendpunct}{\mcitedefaultseppunct}\relax
\EndOfBibitem
\bibitem[Chen \latin{et~al.}(1988)Chen, Gil, Lefebvre, and
  Mathieu]{chen1988exchange}
Chen,~Y.; Gil,~B.; Lefebvre,~P.; Mathieu,~H. Exchange effects on excitons in
  quantum wells. \emph{Phys. Rev. B} \textbf{1988}, \emph{37}, 6429\relax
\mciteBstWouldAddEndPuncttrue
\mciteSetBstMidEndSepPunct{\mcitedefaultmidpunct}
{\mcitedefaultendpunct}{\mcitedefaultseppunct}\relax
\EndOfBibitem
\bibitem[Bester \latin{et~al.}(2003)Bester, Nair, and
  Zunger]{bester2003pseudopotential}
Bester,~G.; Nair,~S.; Zunger,~A. Pseudopotential calculation of the excitonic
  fine structure of million-atom self-assembled In 1- x Ga x A s/G a A s
  quantum dots. \emph{Phys. Rev. B} \textbf{2003}, \emph{67}, 161306\relax
\mciteBstWouldAddEndPuncttrue
\mciteSetBstMidEndSepPunct{\mcitedefaultmidpunct}
{\mcitedefaultendpunct}{\mcitedefaultseppunct}\relax
\EndOfBibitem
\bibitem[Mrowi{\'n}ski \latin{et~al.}(2015)Mrowi{\'n}ski, Musia{l},
  Mary{\'n}ski, Syperek, Misiewicz, Somers, Reithmaier, Hofling, and
  Sek]{mrowinski2015magnetic}
Mrowi{\'n}ski,~P.; Musia{l},~A.; Mary{\'n}ski,~A.; Syperek,~M.; Misiewicz,~J.;
  Somers,~A.; Reithmaier,~J.; Hofling,~S.; Sek,~G. Magnetic field control of
  the neutral and charged exciton fine structure in single quantum dashes
  emitting at 1.55 $\mu$ m. \emph{Appl. Phys. Lett.} \textbf{2015}, \emph{106},
  053114\relax
\mciteBstWouldAddEndPuncttrue
\mciteSetBstMidEndSepPunct{\mcitedefaultmidpunct}
{\mcitedefaultendpunct}{\mcitedefaultseppunct}\relax
\EndOfBibitem
\bibitem[Mrowi{\'n}ski \latin{et~al.}(2016)Mrowi{\'n}ski, Zieli{\'n}ski,
  {\'S}widerski, Misiewicz, Somers, Reithmaier, H{\"o}fling, and
  Sęk]{mrowinski2016excitonic}
Mrowi{\'n}ski,~P.; Zieli{\'n}ski,~M.; {\'S}widerski,~M.; Misiewicz,~J.;
  Somers,~A.; Reithmaier,~J.; H{\"o}fling,~S.; Sęk,~G. Excitonic fine
  structure and binding energies of excitonic complexes in single InAs quantum
  dashes. \emph{Phys. Rev. B} \textbf{2016}, \emph{94}, 115434\relax
\mciteBstWouldAddEndPuncttrue
\mciteSetBstMidEndSepPunct{\mcitedefaultmidpunct}
{\mcitedefaultendpunct}{\mcitedefaultseppunct}\relax
\EndOfBibitem
\bibitem[Li \latin{et~al.}(2003)Li, Wu, Steel, Gammon, Stievater, Katzer, Park,
  Piermarocchi, and Sham]{li2003all}
Li,~X.; Wu,~Y.; Steel,~D.; Gammon,~D.; Stievater,~T.; Katzer,~D.; Park,~D.;
  Piermarocchi,~C.; Sham,~L. An all-optical quantum gate in a semiconductor
  quantum dot. \emph{Science} \textbf{2003}, \emph{301}, 809--811\relax
\mciteBstWouldAddEndPuncttrue
\mciteSetBstMidEndSepPunct{\mcitedefaultmidpunct}
{\mcitedefaultendpunct}{\mcitedefaultseppunct}\relax
\EndOfBibitem
\bibitem[Bonadeo \latin{et~al.}(1998)Bonadeo, Erland, Gammon, Park, Katzer, and
  Steel]{bonadeo1998coherent}
Bonadeo,~N.~H.; Erland,~J.; Gammon,~D.; Park,~D.; Katzer,~D.; Steel,~D.
  Coherent optical control of the quantum state of a single quantum dot.
  \emph{Science} \textbf{1998}, \emph{282}, 1473--1476\relax
\mciteBstWouldAddEndPuncttrue
\mciteSetBstMidEndSepPunct{\mcitedefaultmidpunct}
{\mcitedefaultendpunct}{\mcitedefaultseppunct}\relax
\EndOfBibitem
\bibitem[Stevenson \latin{et~al.}(2006)Stevenson, Young, Atkinson, Cooper,
  Ritchie, and Shields]{stevenson2006semiconductor}
Stevenson,~R.~M.; Young,~R.~J.; Atkinson,~P.; Cooper,~K.; Ritchie,~D.~A.;
  Shields,~A.~J. A semiconductor source of triggered entangled photon pairs.
  \emph{Nature} \textbf{2006}, \emph{439}, 179--182\relax
\mciteBstWouldAddEndPuncttrue
\mciteSetBstMidEndSepPunct{\mcitedefaultmidpunct}
{\mcitedefaultendpunct}{\mcitedefaultseppunct}\relax
\EndOfBibitem
\bibitem[Bennett \latin{et~al.}(2010)Bennett, Pooley, Stevenson, Ward, Patel,
  de~La~Giroday, Sk{\"o}ld, Farrer, Nicoll, Ritchie, \latin{et~al.}
  others]{bennett2010electric}
Bennett,~A.; Pooley,~M.; Stevenson,~R.; Ward,~M.; Patel,~R.;
  de~La~Giroday,~A.~B.; Sk{\"o}ld,~N.; Farrer,~I.; Nicoll,~C.; Ritchie,~D.
  \latin{et~al.}  Electric-field-induced coherent coupling of the exciton
  states in a single quantum dot. \emph{Nat. Phys.} \textbf{2010}, \emph{6},
  947--950\relax
\mciteBstWouldAddEndPuncttrue
\mciteSetBstMidEndSepPunct{\mcitedefaultmidpunct}
{\mcitedefaultendpunct}{\mcitedefaultseppunct}\relax
\EndOfBibitem
\bibitem[Benson \latin{et~al.}(2000)Benson, Santori, Pelton, and
  Yamamoto]{benson2000regulated}
Benson,~O.; Santori,~C.; Pelton,~M.; Yamamoto,~Y. Regulated and entangled
  photons from a single quantum dot. \emph{Phys. Rev. Lett.} \textbf{2000},
  \emph{84}, 2513\relax
\mciteBstWouldAddEndPuncttrue
\mciteSetBstMidEndSepPunct{\mcitedefaultmidpunct}
{\mcitedefaultendpunct}{\mcitedefaultseppunct}\relax
\EndOfBibitem
\bibitem[M{\"u}ller \latin{et~al.}(2014)M{\"u}ller, Bounouar, J{\"o}ns,
  Gl{\"a}ssl, and Michler]{muller2014demand}
M{\"u}ller,~M.; Bounouar,~S.; J{\"o}ns,~K.~D.; Gl{\"a}ssl,~M.; Michler,~P.
  On-demand generation of indistinguishable polarization-entangled photon
  pairs. \emph{Nat. Photonics} \textbf{2014}, \emph{8}, 224--228\relax
\mciteBstWouldAddEndPuncttrue
\mciteSetBstMidEndSepPunct{\mcitedefaultmidpunct}
{\mcitedefaultendpunct}{\mcitedefaultseppunct}\relax
\EndOfBibitem
\bibitem[Dousse \latin{et~al.}(2010)Dousse, Suffczy{\'n}ski, Beveratos, Krebs,
  Lema{\^\i}tre, Sagnes, Bloch, Voisin, and Senellart]{dousse2010ultrabright}
Dousse,~A.; Suffczy{\'n}ski,~J.; Beveratos,~A.; Krebs,~O.; Lema{\^\i}tre,~A.;
  Sagnes,~I.; Bloch,~J.; Voisin,~P.; Senellart,~P. Ultrabright source of
  entangled photon pairs. \emph{Nature} \textbf{2010}, \emph{466},
  217--220\relax
\mciteBstWouldAddEndPuncttrue
\mciteSetBstMidEndSepPunct{\mcitedefaultmidpunct}
{\mcitedefaultendpunct}{\mcitedefaultseppunct}\relax
\EndOfBibitem
\bibitem[Bouwmeester \latin{et~al.}(1997)Bouwmeester, Pan, Mattle, Eibl,
  Weinfurter, and Zeilinger]{bouwmeester1997experimental}
Bouwmeester,~D.; Pan,~J.-W.; Mattle,~K.; Eibl,~M.; Weinfurter,~H.;
  Zeilinger,~A. Experimental quantum teleportation. \emph{Nature}
  \textbf{1997}, \emph{390}, 575--579\relax
\mciteBstWouldAddEndPuncttrue
\mciteSetBstMidEndSepPunct{\mcitedefaultmidpunct}
{\mcitedefaultendpunct}{\mcitedefaultseppunct}\relax
\EndOfBibitem
\bibitem[Briegel \latin{et~al.}(1998)Briegel, D{\"u}r, Cirac, and
  Zoller]{briegel1998quantum}
Briegel,~H.-J.; D{\"u}r,~W.; Cirac,~J.~I.; Zoller,~P. Quantum repeaters: the
  role of imperfect local operations in quantum communication. \emph{Phys. Rev.
  Lett.} \textbf{1998}, \emph{81}, 5932\relax
\mciteBstWouldAddEndPuncttrue
\mciteSetBstMidEndSepPunct{\mcitedefaultmidpunct}
{\mcitedefaultendpunct}{\mcitedefaultseppunct}\relax
\EndOfBibitem
\bibitem[Gisin \latin{et~al.}(2002)Gisin, Ribordy, Tittel, and
  Zbinden]{gisin2002quantum}
Gisin,~N.; Ribordy,~G.; Tittel,~W.; Zbinden,~H. Quantum cryptography.
  \emph{Rev. Mod. Phys.} \textbf{2002}, \emph{74}, 145\relax
\mciteBstWouldAddEndPuncttrue
\mciteSetBstMidEndSepPunct{\mcitedefaultmidpunct}
{\mcitedefaultendpunct}{\mcitedefaultseppunct}\relax
\EndOfBibitem
\bibitem[Fu \latin{et~al.}(1999)Fu, Wang, and Zunger]{fu1999excitonic}
Fu,~H.; Wang,~L.-W.; Zunger,~A. Excitonic exchange splitting in bulk
  semiconductors. \emph{Phys. Rev. B} \textbf{1999}, \emph{59}, 5568\relax
\mciteBstWouldAddEndPuncttrue
\mciteSetBstMidEndSepPunct{\mcitedefaultmidpunct}
{\mcitedefaultendpunct}{\mcitedefaultseppunct}\relax
\EndOfBibitem
\bibitem[Becker \latin{et~al.}(2018)Becker, Vaxenburg, Nedelcu, Sercel,
  Shabaev, Mehl, Michopoulos, Lambrakos, Bernstein, Lyons, \latin{et~al.}
  others]{becker2018bright}
Becker,~M.~A.; Vaxenburg,~R.; Nedelcu,~G.; Sercel,~P.~C.; Shabaev,~A.;
  Mehl,~M.~J.; Michopoulos,~J.~G.; Lambrakos,~S.~G.; Bernstein,~N.;
  Lyons,~J.~L. \latin{et~al.}  Bright triplet excitons in caesium lead halide
  perovskites. \emph{Nature} \textbf{2018}, \emph{553}, 189\relax
\mciteBstWouldAddEndPuncttrue
\mciteSetBstMidEndSepPunct{\mcitedefaultmidpunct}
{\mcitedefaultendpunct}{\mcitedefaultseppunct}\relax
\EndOfBibitem
\bibitem[Isarov \latin{et~al.}(2017)Isarov, Tan, Bodnarchuk, Kovalenko, Rappe,
  and Lifshitz]{isarov2017rashba}
Isarov,~M.; Tan,~L.~Z.; Bodnarchuk,~M.~I.; Kovalenko,~M.~V.; Rappe,~A.~M.;
  Lifshitz,~E. Rashba effect in a single colloidal CsPbBr3 perovskite
  nanocrystal detected by magneto-optical measurements. \emph{Nano Lett.}
  \textbf{2017}, \emph{17}, 5020--5026\relax
\mciteBstWouldAddEndPuncttrue
\mciteSetBstMidEndSepPunct{\mcitedefaultmidpunct}
{\mcitedefaultendpunct}{\mcitedefaultseppunct}\relax
\EndOfBibitem
\bibitem[Sercel \latin{et~al.}(2019)Sercel, Lyons, Wickramaratne, Vaxenburg,
  Bernstein, and Efros]{Sercel2019}
Sercel,~P.~C.; Lyons,~J.~L.; Wickramaratne,~D.; Vaxenburg,~R.; Bernstein,~N.;
  Efros,~A.~L. Exciton Fine Structure in Perovskite Nanocrystals. \emph{Nano
  Lett.} \textbf{2019}, \emph{19}, 4068--4077\relax
\mciteBstWouldAddEndPuncttrue
\mciteSetBstMidEndSepPunct{\mcitedefaultmidpunct}
{\mcitedefaultendpunct}{\mcitedefaultseppunct}\relax
\EndOfBibitem
\bibitem[Fu \latin{et~al.}(2017)Fu, Tamarat, Huang, Even, Rogach, and
  Lounis]{fu2017neutral}
Fu,~M.; Tamarat,~P.; Huang,~H.; Even,~J.; Rogach,~A.~L.; Lounis,~B. Neutral and
  charged exciton fine structure in single lead halide perovskite nanocrystals
  revealed by magneto-optical spectroscopy. \emph{Nano Lett.} \textbf{2017},
  \emph{17}, 2895--2901\relax
\mciteBstWouldAddEndPuncttrue
\mciteSetBstMidEndSepPunct{\mcitedefaultmidpunct}
{\mcitedefaultendpunct}{\mcitedefaultseppunct}\relax
\EndOfBibitem
\bibitem[Ramade \latin{et~al.}(2018)Ramade, Andriambariarijaona, Steinmetz,
  Goubet, Legrand, Barisien, Bernardot, Testelin, Lhuillier, Bramati,
  \latin{et~al.} others]{ramade2018fine}
Ramade,~J.; Andriambariarijaona,~L.~M.; Steinmetz,~V.; Goubet,~N.; Legrand,~L.;
  Barisien,~T.; Bernardot,~F.; Testelin,~C.; Lhuillier,~E.; Bramati,~A.
  \latin{et~al.}  Fine structure of excitons and electron--hole exchange energy
  in polymorphic CsPbBr 3 single nanocrystals. \emph{Nanoscale} \textbf{2018},
  \emph{10}, 6393--6401\relax
\mciteBstWouldAddEndPuncttrue
\mciteSetBstMidEndSepPunct{\mcitedefaultmidpunct}
{\mcitedefaultendpunct}{\mcitedefaultseppunct}\relax
\EndOfBibitem
\bibitem[Aich \latin{et~al.}(2019)Aich, Sa{\"\i}di, Radhia, Boujdaria,
  Barisien, Legrand, Bernardot, Chamarro, and Testelin]{aich2019bright}
Aich,~R.~B.; Sa{\"\i}di,~I.; Radhia,~S.~B.; Boujdaria,~K.; Barisien,~T.;
  Legrand,~L.; Bernardot,~F.; Chamarro,~M.; Testelin,~C. Bright-Exciton
  Splittings in Inorganic Cesium Lead Halide Perovskite Nanocrystals.
  \emph{Phys. Rev. Appl.} \textbf{2019}, \emph{11}, 034042\relax
\mciteBstWouldAddEndPuncttrue
\mciteSetBstMidEndSepPunct{\mcitedefaultmidpunct}
{\mcitedefaultendpunct}{\mcitedefaultseppunct}\relax
\EndOfBibitem
\bibitem[Tamarat \latin{et~al.}(2019)Tamarat, Bodnarchuk, Trebbia, Erni,
  Kovalenko, Even, and Lounis]{Tamarat2019}
Tamarat,~P.; Bodnarchuk,~M.~I.; Trebbia,~J.-B.; Erni,~R.; Kovalenko,~M.~V.;
  Even,~J.; Lounis,~B. The ground exciton state of formamidinium lead bromide
  perovskite nanocrystals is a singlet dark state. \emph{Nat. Mater.}
  \textbf{2019}, \emph{18}, 717--724\relax
\mciteBstWouldAddEndPuncttrue
\mciteSetBstMidEndSepPunct{\mcitedefaultmidpunct}
{\mcitedefaultendpunct}{\mcitedefaultseppunct}\relax
\EndOfBibitem
\bibitem[Even \latin{et~al.}(2013)Even, Pedesseau, Jancu, and
  Katan]{even2013importance}
Even,~J.; Pedesseau,~L.; Jancu,~J.-M.; Katan,~C. Importance of spin--orbit
  coupling in hybrid organic/inorganic perovskites for photovoltaic
  applications. \emph{J. Phys. Chem. Lett.} \textbf{2013}, \emph{4},
  2999--3005\relax
\mciteBstWouldAddEndPuncttrue
\mciteSetBstMidEndSepPunct{\mcitedefaultmidpunct}
{\mcitedefaultendpunct}{\mcitedefaultseppunct}\relax
\EndOfBibitem
\bibitem[Even \latin{et~al.}(2014)Even, Pedesseau, Jancu, and
  Katan]{even2014dft}
Even,~J.; Pedesseau,~L.; Jancu,~J.-M.; Katan,~C. DFT and k{\textperiodcentered}
  p modelling of the phase transitions of lead and tin halide perovskites for
  photovoltaic cells. \emph{Phys. Status Solidi RRL} \textbf{2014}, \emph{8},
  31--35\relax
\mciteBstWouldAddEndPuncttrue
\mciteSetBstMidEndSepPunct{\mcitedefaultmidpunct}
{\mcitedefaultendpunct}{\mcitedefaultseppunct}\relax
\EndOfBibitem
\bibitem[Brivio \latin{et~al.}(2014)Brivio, Butler, Walsh, and
  Van~Schilfgaarde]{brivio2014relativistic}
Brivio,~F.; Butler,~K.~T.; Walsh,~A.; Van~Schilfgaarde,~M. Relativistic
  quasiparticle self-consistent electronic structure of hybrid halide
  perovskite photovoltaic absorbers. \emph{Phys. Rev. B} \textbf{2014},
  \emph{89}, 155204\relax
\mciteBstWouldAddEndPuncttrue
\mciteSetBstMidEndSepPunct{\mcitedefaultmidpunct}
{\mcitedefaultendpunct}{\mcitedefaultseppunct}\relax
\EndOfBibitem
\bibitem[Endres \latin{et~al.}(2016)Endres, Egger, Kulbak, Kerner, Zhao,
  Silver, Hodes, Rand, Cahen, Kronik, \latin{et~al.} others]{endres2016valence}
Endres,~J.; Egger,~D.~A.; Kulbak,~M.; Kerner,~R.~A.; Zhao,~L.; Silver,~S.~H.;
  Hodes,~G.; Rand,~B.~P.; Cahen,~D.; Kronik,~L. \latin{et~al.}  Valence and
  Conduction Band Densities of States of Metal Halide Perovskites: A Combined
  Experimental--Theoretical Study. \emph{J. Phys. Chem. Lett.} \textbf{2016},
  \emph{7}, 2722--2729\relax
\mciteBstWouldAddEndPuncttrue
\mciteSetBstMidEndSepPunct{\mcitedefaultmidpunct}
{\mcitedefaultendpunct}{\mcitedefaultseppunct}\relax
\EndOfBibitem
\bibitem[Poglitsch and Weber(1987)Poglitsch, and Weber]{poglitsch1987dynamic}
Poglitsch,~A.; Weber,~D. Dynamic disorder in methylammoniumtrihalogenoplumbates
  (II) observed by millimeter-wave spectroscopy. \emph{J. Chem. Phys.}
  \textbf{1987}, \emph{87}, 6373--6378\relax
\mciteBstWouldAddEndPuncttrue
\mciteSetBstMidEndSepPunct{\mcitedefaultmidpunct}
{\mcitedefaultendpunct}{\mcitedefaultseppunct}\relax
\EndOfBibitem
\bibitem[Chen \latin{et~al.}(2018)Chen, Hu, Lu, Chang, Shi, Li, Zhong, and
  Han]{chen2018elucidating}
Chen,~C.; Hu,~X.; Lu,~W.; Chang,~S.; Shi,~L.; Li,~L.; Zhong,~H.; Han,~J.-B.
  Elucidating the phase transitions and temperature-dependent photoluminescence
  of MAPbBr3 single crystal. \emph{J. Phys. D: Appl. Phys.} \textbf{2018},
  \emph{51}, 045105\relax
\mciteBstWouldAddEndPuncttrue
\mciteSetBstMidEndSepPunct{\mcitedefaultmidpunct}
{\mcitedefaultendpunct}{\mcitedefaultseppunct}\relax
\EndOfBibitem
\bibitem[Odenthal \latin{et~al.}(2017)Odenthal, Talmadge, Gundlach, Wang,
  Zhang, Sun, Yu, Vardeny, and Li]{odenthal2017spin}
Odenthal,~P.; Talmadge,~W.; Gundlach,~N.; Wang,~R.; Zhang,~C.; Sun,~D.;
  Yu,~Z.-G.; Vardeny,~Z.~V.; Li,~Y.~S. Spin-polarized exciton quantum beating
  in hybrid organic-inorganic perovskites. \emph{Nat. Phys.} \textbf{2017},
  \relax
\mciteBstWouldAddEndPunctfalse
\mciteSetBstMidEndSepPunct{\mcitedefaultmidpunct}
{}{\mcitedefaultseppunct}\relax
\EndOfBibitem
\bibitem[Yu(2016)]{yu2016effective}
Yu,~Z. Effective-mass model and magneto-optical properties in hybrid
  perovskites. \emph{Sci. Rep.} \textbf{2016}, \emph{6}, 28576\relax
\mciteBstWouldAddEndPuncttrue
\mciteSetBstMidEndSepPunct{\mcitedefaultmidpunct}
{\mcitedefaultendpunct}{\mcitedefaultseppunct}\relax
\EndOfBibitem
\bibitem[Nestoklon \latin{et~al.}(2018)Nestoklon, Goupalov, Dzhioev, Ken,
  Korenev, Kusrayev, Sapega, de~Weerd, Gomez, Gregorkiewicz, \latin{et~al.}
  others]{nestoklon2018optical}
Nestoklon,~M.; Goupalov,~S.; Dzhioev,~R.; Ken,~O.; Korenev,~V.;
  Kusrayev,~Y.~G.; Sapega,~V.; de~Weerd,~C.; Gomez,~L.; Gregorkiewicz,~T.
  \latin{et~al.}  Optical orientation and alignment of excitons in ensembles of
  inorganic perovskite nanocrystals. \emph{Phys. Rev. B} \textbf{2018},
  \emph{97}, 235304\relax
\mciteBstWouldAddEndPuncttrue
\mciteSetBstMidEndSepPunct{\mcitedefaultmidpunct}
{\mcitedefaultendpunct}{\mcitedefaultseppunct}\relax
\EndOfBibitem
\bibitem[Yin \latin{et~al.}(2017)Yin, Chen, Song, Lv, Hu, Sun, William, Zhang,
  Wang, Zhang, \latin{et~al.} others]{yin2017bright}
Yin,~C.; Chen,~L.; Song,~N.; Lv,~Y.; Hu,~F.; Sun,~C.; William,~W.~Y.;
  Zhang,~C.; Wang,~X.; Zhang,~Y. \latin{et~al.}  Bright-Exciton Fine-Structure
  Splittings in Single Perovskite Nanocrystals. \emph{Phys. Rev. Lett.}
  \textbf{2017}, \emph{119}, 026401\relax
\mciteBstWouldAddEndPuncttrue
\mciteSetBstMidEndSepPunct{\mcitedefaultmidpunct}
{\mcitedefaultendpunct}{\mcitedefaultseppunct}\relax
\EndOfBibitem
\bibitem[Pfingsten \latin{et~al.}(2018)Pfingsten, Klein, Protesescu,
  Bodnarchuk, Kovalenko, and Bacher]{pfingsten2018phonon}
Pfingsten,~O.; Klein,~J.; Protesescu,~L.; Bodnarchuk,~M.~I.; Kovalenko,~M.~V.;
  Bacher,~G. Phonon Interaction and Phase Transition in Single Formamidinium
  Lead Bromide Quantum Dots. \emph{Nano Lett.} \textbf{2018}, \emph{18},
  4440--4446\relax
\mciteBstWouldAddEndPuncttrue
\mciteSetBstMidEndSepPunct{\mcitedefaultmidpunct}
{\mcitedefaultendpunct}{\mcitedefaultseppunct}\relax
\EndOfBibitem
\bibitem[Rainò \latin{et~al.}(2016)Rainò, Nedelcu, Protesescu, Bodnarchuk,
  Kovalenko, Mahrt, and Stöferle]{raino2016single}
Rainò,~G.; Nedelcu,~G.; Protesescu,~L.; Bodnarchuk,~M.~I.; Kovalenko,~M.~V.;
  Mahrt,~R.~F.; Stöferle,~T. Single cesium lead halide perovskite
  nanocrystals at low temperature: fast single-photon emission, reduced
  blinking, and exciton fine structure. \emph{ACS nano} \textbf{2016},
  \emph{10}, 2485--2490\relax
\mciteBstWouldAddEndPuncttrue
\mciteSetBstMidEndSepPunct{\mcitedefaultmidpunct}
{\mcitedefaultendpunct}{\mcitedefaultseppunct}\relax
\EndOfBibitem
\bibitem[Ramade \latin{et~al.}(2018)Ramade, Andriambariarijaona, Steinmetz,
  Goubet, Legrand, Barisien, Bernardot, Testelin, Lhuillier, Bramati,
  \latin{et~al.} others]{ramade2018exciton}
Ramade,~J.; Andriambariarijaona,~L.~M.; Steinmetz,~V.; Goubet,~N.; Legrand,~L.;
  Barisien,~T.; Bernardot,~F.; Testelin,~C.; Lhuillier,~E.; Bramati,~A.
  \latin{et~al.}  Exciton-phonon coupling in a CsPbBr3 single nanocrystal.
  \emph{Appl. Phys. Lett.} \textbf{2018}, \emph{112}, 072104\relax
\mciteBstWouldAddEndPuncttrue
\mciteSetBstMidEndSepPunct{\mcitedefaultmidpunct}
{\mcitedefaultendpunct}{\mcitedefaultseppunct}\relax
\EndOfBibitem
\bibitem[Zieli{\'n}ski(2013)]{zielinski2013excitonic}
Zieli{\'n}ski,~M. Excitonic fine structure of elongated InAs/InP quantum dots.
  \emph{Phys. Rev. B} \textbf{2013}, \emph{88}, 155319\relax
\mciteBstWouldAddEndPuncttrue
\mciteSetBstMidEndSepPunct{\mcitedefaultmidpunct}
{\mcitedefaultendpunct}{\mcitedefaultseppunct}\relax
\EndOfBibitem
\bibitem[Seguin \latin{et~al.}(2005)Seguin, Schliwa, Rodt, P{\"o}tschke, Pohl,
  and Bimberg]{seguin2005size}
Seguin,~R.; Schliwa,~A.; Rodt,~S.; P{\"o}tschke,~K.; Pohl,~U.; Bimberg,~D.
  Size-dependent fine-structure splitting in self-organized InAs/GaAs quantum
  dots. \emph{Phys. Rev. Lett.} \textbf{2005}, \emph{95}, 257402\relax
\mciteBstWouldAddEndPuncttrue
\mciteSetBstMidEndSepPunct{\mcitedefaultmidpunct}
{\mcitedefaultendpunct}{\mcitedefaultseppunct}\relax
\EndOfBibitem
\bibitem[Ben~Aich \latin{et~al.}()Ben~Aich, Saidi, Ben~Radhia, Boujdaria,
  Barisien, Legrand, Bernandot, Chamarro, and Testelin]{inprep2}
Ben~Aich,~R.; Saidi,~I.; Ben~Radhia,~S.; Boujdaria,~K.; Barisien,~T.;
  Legrand,~T.; Bernandot,~F.; Chamarro,~M.; Testelin,~C. Article in
  preparation. \emph{Submitted to Phys. Rev. Appl.} \relax
\mciteBstWouldAddEndPunctfalse
\mciteSetBstMidEndSepPunct{\mcitedefaultmidpunct}
{}{\mcitedefaultseppunct}\relax
\EndOfBibitem
\bibitem[Niesner \latin{et~al.}(2016)Niesner, Wilhelm, Levchuk, Osvet,
  Shrestha, Batentschuk, Brabec, and Fauster]{niesner2016giant}
Niesner,~D.; Wilhelm,~M.; Levchuk,~I.; Osvet,~A.; Shrestha,~S.;
  Batentschuk,~M.; Brabec,~C.; Fauster,~T. Giant Rashba Splitting in CH 3 NH 3
  PbBr 3 Organic-Inorganic Perovskite. \emph{Phys. Rev. Lett.} \textbf{2016},
  \emph{117}, 126401\relax
\mciteBstWouldAddEndPuncttrue
\mciteSetBstMidEndSepPunct{\mcitedefaultmidpunct}
{\mcitedefaultendpunct}{\mcitedefaultseppunct}\relax
\EndOfBibitem
\bibitem[Mosconi \latin{et~al.}(2017)Mosconi, Etienne, and
  De~Angelis]{mosconi2017rashba}
Mosconi,~E.; Etienne,~T.; De~Angelis,~F. Rashba Band Splitting in Organohalide
  Lead Perovskites: Bulk and Surface Effects. \emph{J. Phys. Chem. Lett.}
  \textbf{2017}, \relax
\mciteBstWouldAddEndPunctfalse
\mciteSetBstMidEndSepPunct{\mcitedefaultmidpunct}
{}{\mcitedefaultseppunct}\relax
\EndOfBibitem
\bibitem[Swarnkar \latin{et~al.}(2016)Swarnkar, Marshall, Sanehira,
  Chernomordik, Moore, Christians, Chakrabarti, and
  Luther]{swarnkar2016quantum}
Swarnkar,~A.; Marshall,~A.~R.; Sanehira,~E.~M.; Chernomordik,~B.~D.;
  Moore,~D.~T.; Christians,~J.~A.; Chakrabarti,~T.; Luther,~J.~M. Quantum
  dot--induced phase stabilization of $\alpha$-CsPbI3 perovskite for
  high-efficiency photovoltaics. \emph{Science} \textbf{2016}, \emph{354},
  92--95\relax
\mciteBstWouldAddEndPuncttrue
\mciteSetBstMidEndSepPunct{\mcitedefaultmidpunct}
{\mcitedefaultendpunct}{\mcitedefaultseppunct}\relax
\EndOfBibitem
\bibitem[Nayak \latin{et~al.}(2016)Nayak, Moore, Wenger, Nayak, Haghighirad,
  Fineberg, Noel, Reid, Rumbles, Kukura, \latin{et~al.}
  others]{nayak2016mechanism}
Nayak,~P.~K.; Moore,~D.~T.; Wenger,~B.; Nayak,~S.; Haghighirad,~A.~A.;
  Fineberg,~A.; Noel,~N.~K.; Reid,~O.~G.; Rumbles,~G.; Kukura,~P.
  \latin{et~al.}  Mechanism for rapid growth of organic--inorganic halide
  perovskite crystals. \emph{Nat. Commun.} \textbf{2016}, \emph{7}, 13303\relax
\mciteBstWouldAddEndPuncttrue
\mciteSetBstMidEndSepPunct{\mcitedefaultmidpunct}
{\mcitedefaultendpunct}{\mcitedefaultseppunct}\relax
\EndOfBibitem
\bibitem[Tilchin \latin{et~al.}(2016)Tilchin, Dirin, Maikov, Sashchiuk,
  Kovalenko, and Lifshitz]{tilchin2016hydrogen}
Tilchin,~J.; Dirin,~D.~N.; Maikov,~G.~I.; Sashchiuk,~A.; Kovalenko,~M.~V.;
  Lifshitz,~E. Hydrogen-like Wannier--Mott Excitons in Single Crystal of
  Methylammonium Lead Bromide Perovskite. \emph{ACS nano} \textbf{2016},
  \emph{10}, 6363--6371\relax
\mciteBstWouldAddEndPuncttrue
\mciteSetBstMidEndSepPunct{\mcitedefaultmidpunct}
{\mcitedefaultendpunct}{\mcitedefaultseppunct}\relax
\EndOfBibitem
\bibitem[Wright \latin{et~al.}(2016)Wright, Verdi, Milot, Eperon,
  P{\'e}rez-Osorio, Snaith, Giustino, Johnston, and Herz]{wright2016electron}
Wright,~A.~D.; Verdi,~C.; Milot,~R.~L.; Eperon,~G.~E.; P{\'e}rez-Osorio,~M.~A.;
  Snaith,~H.~J.; Giustino,~F.; Johnston,~M.~B.; Herz,~L.~M. Electron--phonon
  coupling in hybrid lead halide perovskites. \emph{Nat. Commun.}
  \textbf{2016}, \emph{7}, 11755\relax
\mciteBstWouldAddEndPuncttrue
\mciteSetBstMidEndSepPunct{\mcitedefaultmidpunct}
{\mcitedefaultendpunct}{\mcitedefaultseppunct}\relax
\EndOfBibitem
\bibitem[Kataoka \latin{et~al.}(1993)Kataoka, Kondo, Ito, Sasaki, Uchida, and
  Miura]{kataoka1993magneto}
Kataoka,~T.; Kondo,~T.; Ito,~R.; Sasaki,~S.; Uchida,~K.; Miura,~N.
  Magneto-optical study on excitonic spectra in (C 6 H 13 NH 3) 2 PbI 4.
  \emph{Phys. Rev. B} \textbf{1993}, \emph{47}, 2010\relax
\mciteBstWouldAddEndPuncttrue
\mciteSetBstMidEndSepPunct{\mcitedefaultmidpunct}
{\mcitedefaultendpunct}{\mcitedefaultseppunct}\relax
\EndOfBibitem
\bibitem[Tanaka \latin{et~al.}(2005)Tanaka, Takahashi, Kondo, Umeda, Ema,
  Umebayashi, Asai, Uchida, and Miura]{tanaka2005electronic}
Tanaka,~K.; Takahashi,~T.; Kondo,~T.; Umeda,~K.; Ema,~K.; Umebayashi,~T.;
  Asai,~K.; Uchida,~K.; Miura,~N. Electronic and excitonic structures of
  inorganic--organic perovskite-type quantum-well crystal (C4H9NH3) 2PbBr4.
  \emph{Jpn. J. Appl. Phys.} \textbf{2005}, \emph{44}, 5923\relax
\mciteBstWouldAddEndPuncttrue
\mciteSetBstMidEndSepPunct{\mcitedefaultmidpunct}
{\mcitedefaultendpunct}{\mcitedefaultseppunct}\relax
\EndOfBibitem
\bibitem[Ndione \latin{et~al.}(2016)Ndione, Li, and Zhu]{ndione2016effects}
Ndione,~P.~F.; Li,~Z.; Zhu,~K. Effects of alloying on the optical properties of
  organic--inorganic lead halide perovskite thin films. \emph{Journal of
  Materials Chemistry C} \textbf{2016}, \emph{4}, 7775--7782\relax
\mciteBstWouldAddEndPuncttrue
\mciteSetBstMidEndSepPunct{\mcitedefaultmidpunct}
{\mcitedefaultendpunct}{\mcitedefaultseppunct}\relax
\EndOfBibitem
\bibitem[Valverde-Ch{\'a}vez \latin{et~al.}(2015)Valverde-Ch{\'a}vez, Ponseca,
  Stoumpos, Yartsev, Kanatzidis, Sundstr{\"o}m, and
  Cooke]{valverde2015intrinsic}
Valverde-Ch{\'a}vez,~D.~A.; Ponseca,~C.~S.; Stoumpos,~C.~C.; Yartsev,~A.;
  Kanatzidis,~M.~G.; Sundstr{\"o}m,~V.; Cooke,~D.~G. Intrinsic femtosecond
  charge generation dynamics in single crystal CH 3 NH 3 PbI 3. \emph{Energy
  Environ. Sci.} \textbf{2015}, \emph{8}, 3700--3707\relax
\mciteBstWouldAddEndPuncttrue
\mciteSetBstMidEndSepPunct{\mcitedefaultmidpunct}
{\mcitedefaultendpunct}{\mcitedefaultseppunct}\relax
\EndOfBibitem
\bibitem[Zhao \latin{et~al.}(2017)Zhao, Skelton, Hu, La-o Vorakiat, Zhu,
  Marcus, Michel-Beyerle, Lam, Walsh, and Chia]{zhao2017low}
Zhao,~D.; Skelton,~J.~M.; Hu,~H.; La-o Vorakiat,~C.; Zhu,~J.-X.; Marcus,~R.~A.;
  Michel-Beyerle,~M.-E.; Lam,~Y.~M.; Walsh,~A.; Chia,~E.~E. Low-frequency
  optical phonon modes and carrier mobility in the halide perovskite
  CH3NH3PbBr3 using terahertz time-domain spectroscopy. \emph{Appl. Phys.
  Lett.} \textbf{2017}, \emph{111}, 201903\relax
\mciteBstWouldAddEndPuncttrue
\mciteSetBstMidEndSepPunct{\mcitedefaultmidpunct}
{\mcitedefaultendpunct}{\mcitedefaultseppunct}\relax
\EndOfBibitem
\bibitem[Anusca \latin{et~al.}(2017)Anusca, Bal{\v{c}}i{\=u}nas, Gemeiner,
  Svirskas, Sanlialp, Lackner, Fettkenhauer, Belovickis, Samulionis, Ivanov,
  \latin{et~al.} others]{anusca2017dielectric}
Anusca,~I.; Bal{\v{c}}i{\=u}nas,~S.; Gemeiner,~P.; Svirskas,~{\v{S}}.;
  Sanlialp,~M.; Lackner,~G.; Fettkenhauer,~C.; Belovickis,~J.; Samulionis,~V.;
  Ivanov,~M. \latin{et~al.}  Dielectric response: Answer to many questions in
  the methylammonium lead halide solar cell absorbers. \emph{Adv. Energy
  Mater.} \textbf{2017}, \emph{7}, 1700600\relax
\mciteBstWouldAddEndPuncttrue
\mciteSetBstMidEndSepPunct{\mcitedefaultmidpunct}
{\mcitedefaultendpunct}{\mcitedefaultseppunct}\relax
\EndOfBibitem
\bibitem[Miura(2008)]{miura2008physics}
Miura,~N. \emph{Physics of semiconductors in high magnetic fields}; Oxford
  University Press: New York, 2008; Vol.~15\relax
\mciteBstWouldAddEndPuncttrue
\mciteSetBstMidEndSepPunct{\mcitedefaultmidpunct}
{\mcitedefaultendpunct}{\mcitedefaultseppunct}\relax
\EndOfBibitem
\bibitem[Galkowski \latin{et~al.}(2016)Galkowski, Mitioglu, Miyata, Plochocka,
  Portugall, Eperon, Wang, Stergiopoulos, Stranks, Snaith, \latin{et~al.}
  others]{galkowski2016determination}
Galkowski,~K.; Mitioglu,~A.; Miyata,~A.; Plochocka,~P.; Portugall,~O.;
  Eperon,~G.~E.; Wang,~J. T.-W.; Stergiopoulos,~T.; Stranks,~S.~D.;
  Snaith,~H.~J. \latin{et~al.}  Determination of the exciton binding energy and
  effective masses for methylammonium and formamidinium lead tri-halide
  perovskite semiconductors. \emph{Energy Environ. Sci.} \textbf{2016},
  \emph{9}, 962--970\relax
\mciteBstWouldAddEndPuncttrue
\mciteSetBstMidEndSepPunct{\mcitedefaultmidpunct}
{\mcitedefaultendpunct}{\mcitedefaultseppunct}\relax
\EndOfBibitem
\bibitem[Yang \latin{et~al.}(2017)Yang, Surrente, Galkowski, Miyata, Portugall,
  Sutton, Haghighirad, Snaith, Maude, Plochocka, \latin{et~al.}
  others]{yang2017impact}
Yang,~Z.; Surrente,~A.; Galkowski,~K.; Miyata,~A.; Portugall,~O.;
  Sutton,~R.~J.; Haghighirad,~A.; Snaith,~H.~J.; Maude,~D.~K.; Plochocka,~P.
  \latin{et~al.}  Impact of the halide cage on the electronic properties of
  fully inorganic cesium lead halide perovskites. \emph{ACS Energy Lett.}
  \textbf{2017}, \emph{2}, 1621--1627\relax
\mciteBstWouldAddEndPuncttrue
\mciteSetBstMidEndSepPunct{\mcitedefaultmidpunct}
{\mcitedefaultendpunct}{\mcitedefaultseppunct}\relax
\EndOfBibitem
\bibitem[Kepenekian and Even(2017)Kepenekian, and Even]{kepenekian2017rashba}
Kepenekian,~M.; Even,~J. Rashba and Dresselhaus Couplings in Halide
  Perovskites: Accomplishments and Opportunities for Spintronics and
  Spin--Orbitronics. \emph{J. Phys. Chem. Lett.} \textbf{2017}, \emph{8},
  3362--3370\relax
\mciteBstWouldAddEndPuncttrue
\mciteSetBstMidEndSepPunct{\mcitedefaultmidpunct}
{\mcitedefaultendpunct}{\mcitedefaultseppunct}\relax
\EndOfBibitem
\bibitem[Zheng \latin{et~al.}(2015)Zheng, Tan, Liu, and Rappe]{zheng2015rashba}
Zheng,~F.; Tan,~L.~Z.; Liu,~S.; Rappe,~A.~M. Rashba spin--orbit coupling
  enhanced carrier lifetime in CH3NH3PbI3. \emph{Nano Lett.} \textbf{2015},
  \emph{15}, 7794--7800\relax
\mciteBstWouldAddEndPuncttrue
\mciteSetBstMidEndSepPunct{\mcitedefaultmidpunct}
{\mcitedefaultendpunct}{\mcitedefaultseppunct}\relax
\EndOfBibitem
\bibitem[Etienne \latin{et~al.}(2016)Etienne, Mosconi, and
  De~Angelis]{etienne2016dynamical}
Etienne,~T.; Mosconi,~E.; De~Angelis,~F. Dynamical Origin of the Rashba Effect
  in Organohalide Lead Perovskites: A Key to Suppressed Carrier Recombination
  in Perovskite Solar Cells? \emph{J. Phys. Chem. Lett.} \textbf{2016},
  \emph{7}, 1638--1645\relax
\mciteBstWouldAddEndPuncttrue
\mciteSetBstMidEndSepPunct{\mcitedefaultmidpunct}
{\mcitedefaultendpunct}{\mcitedefaultseppunct}\relax
\EndOfBibitem
\bibitem[Quarti \latin{et~al.}(2014)Quarti, Mosconi, and
  De~Angelis]{quarti2014interplay}
Quarti,~C.; Mosconi,~E.; De~Angelis,~F. Interplay of orientational order and
  electronic structure in methylammonium lead iodide: implications for solar
  cell operation. \emph{Chem. Mater.} \textbf{2014}, \emph{26},
  6557--6569\relax
\mciteBstWouldAddEndPuncttrue
\mciteSetBstMidEndSepPunct{\mcitedefaultmidpunct}
{\mcitedefaultendpunct}{\mcitedefaultseppunct}\relax
\EndOfBibitem
\bibitem[Yaffe \latin{et~al.}(2017)Yaffe, Guo, Tan, Egger, Hull, Stoumpos,
  Zheng, Heinz, Kronik, Kanatzidis, \latin{et~al.} others]{yaffe2017local}
Yaffe,~O.; Guo,~Y.; Tan,~L.~Z.; Egger,~D.~A.; Hull,~T.; Stoumpos,~C.~C.;
  Zheng,~F.; Heinz,~T.~F.; Kronik,~L.; Kanatzidis,~M.~G. \latin{et~al.}  Local
  polar fluctuations in lead halide perovskite crystals. \emph{Phys. Rev.
  Lett.} \textbf{2017}, \emph{118}, 136001\relax
\mciteBstWouldAddEndPuncttrue
\mciteSetBstMidEndSepPunct{\mcitedefaultmidpunct}
{\mcitedefaultendpunct}{\mcitedefaultseppunct}\relax
\EndOfBibitem
\bibitem[Dar \latin{et~al.}(2016)Dar, Jacopin, Meloni, Mattoni, Arora, Boziki,
  Zakeeruddin, Rothlisberger, and Gr{\"a}tzel]{dar2016origin}
Dar,~M.~I.; Jacopin,~G.; Meloni,~S.; Mattoni,~A.; Arora,~N.; Boziki,~A.;
  Zakeeruddin,~S.~M.; Rothlisberger,~U.; Gr{\"a}tzel,~M. Origin of unusual
  bandgap shift and dual emission in organic-inorganic lead halide perovskites.
  \emph{Sci. Adv.} \textbf{2016}, \emph{2}, e1601156\relax
\mciteBstWouldAddEndPuncttrue
\mciteSetBstMidEndSepPunct{\mcitedefaultmidpunct}
{\mcitedefaultendpunct}{\mcitedefaultseppunct}\relax
\EndOfBibitem
\bibitem[Niesner \latin{et~al.}(2018)Niesner, Hauck, Shrestha, Levchuk, Matt,
  Osvet, Batentschuk, Brabec, Weber, and Fauster]{niesner2018structural}
Niesner,~D.; Hauck,~M.; Shrestha,~S.; Levchuk,~I.; Matt,~G.~J.; Osvet,~A.;
  Batentschuk,~M.; Brabec,~C.; Weber,~H.~B.; Fauster,~T. Structural
  fluctuations cause spin-split states in tetragonal (CH3NH3) PbI3 as evidenced
  by the circular photogalvanic effect. \emph{Proceedings of the National
  Academy of Sciences} \textbf{2018}, \emph{115}, 9509--9514\relax
\mciteBstWouldAddEndPuncttrue
\mciteSetBstMidEndSepPunct{\mcitedefaultmidpunct}
{\mcitedefaultendpunct}{\mcitedefaultseppunct}\relax
\EndOfBibitem
\bibitem[Yang \latin{et~al.}(2015)Yang, Yan, Yang, Choi, Zhu, Luther, and
  Beard]{yang2015low}
Yang,~Y.; Yan,~Y.; Yang,~M.; Choi,~S.; Zhu,~K.; Luther,~J.~M.; Beard,~M.~C. Low
  surface recombination velocity in solution-grown CH 3 NH 3 PbBr 3 perovskite
  single crystal. \emph{Nat. Commun.} \textbf{2015}, \emph{6}, 7961\relax
\mciteBstWouldAddEndPuncttrue
\mciteSetBstMidEndSepPunct{\mcitedefaultmidpunct}
{\mcitedefaultendpunct}{\mcitedefaultseppunct}\relax
\EndOfBibitem
\bibitem[Sabbah and Riffe(2002)Sabbah, and Riffe]{sabbah2002femtosecond}
Sabbah,~A.; Riffe,~D.~M. Femtosecond pump-probe reflectivity study of silicon
  carrier dynamics. \emph{Phys. Rev. B} \textbf{2002}, \emph{66}, 165217\relax
\mciteBstWouldAddEndPuncttrue
\mciteSetBstMidEndSepPunct{\mcitedefaultmidpunct}
{\mcitedefaultendpunct}{\mcitedefaultseppunct}\relax
\EndOfBibitem
\bibitem[Canneson \latin{et~al.}(2017)Canneson, Shornikova, Yakovlev, Rogge,
  Mitioglu, Ballottin, Christianen, Lhuillier, Bayer, and
  Biadala]{canneson2017negatively}
Canneson,~D.; Shornikova,~E.~V.; Yakovlev,~D.~R.; Rogge,~T.; Mitioglu,~A.~A.;
  Ballottin,~M.~V.; Christianen,~P.~C.; Lhuillier,~E.; Bayer,~M.; Biadala,~L.
  Negatively charged and dark excitons in CsPbBr3 perovskite nanocrystals
  revealed by high magnetic fields. \emph{Nano Lett.} \textbf{2017}, \emph{17},
  6177--6183\relax
\mciteBstWouldAddEndPuncttrue
\mciteSetBstMidEndSepPunct{\mcitedefaultmidpunct}
{\mcitedefaultendpunct}{\mcitedefaultseppunct}\relax
\EndOfBibitem
\bibitem[Xu \latin{et~al.}(2018)Xu, Vliem, and Meijerink]{xu2018long}
Xu,~K.; Vliem,~J.~F.; Meijerink,~A. Long-Lived Dark Exciton Emission in
  Mn-Doped CsPbCl3 Perovskite Nanocrystals. \emph{J. Mater. Chem. C}
  \textbf{2018}, \emph{123}, 979--984\relax
\mciteBstWouldAddEndPuncttrue
\mciteSetBstMidEndSepPunct{\mcitedefaultmidpunct}
{\mcitedefaultendpunct}{\mcitedefaultseppunct}\relax
\EndOfBibitem
\bibitem[Zhu and Podzorov(2015)Zhu, and Podzorov]{zhu2015charge}
Zhu,~X.-Y.; Podzorov,~V. Charge carriers in hybrid organic--inorganic lead
  halide perovskites might be protected as large polarons. \emph{J. Phys. Chem.
  Lett.} \textbf{2015}, \emph{6}, 4758--4761\relax
\mciteBstWouldAddEndPuncttrue
\mciteSetBstMidEndSepPunct{\mcitedefaultmidpunct}
{\mcitedefaultendpunct}{\mcitedefaultseppunct}\relax
\EndOfBibitem
\bibitem[Schlipf \latin{et~al.}(2018)Schlipf, Ponc{\'e}, and
  Giustino]{schlipf2018carrier}
Schlipf,~M.; Ponc{\'e},~S.; Giustino,~F. Carrier lifetimes and polaronic mass
  enhancement in the hybrid halide perovskite CH 3 NH 3 PbI 3 from multiphonon
  Fr{\"o}hlich coupling. \emph{Phys. Rev. Lett.} \textbf{2018}, \emph{121},
  086402\relax
\mciteBstWouldAddEndPuncttrue
\mciteSetBstMidEndSepPunct{\mcitedefaultmidpunct}
{\mcitedefaultendpunct}{\mcitedefaultseppunct}\relax
\EndOfBibitem
\bibitem[Sendner \latin{et~al.}(2016)Sendner, Nayak, Egger, Beck, M{\"u}ller,
  Epding, Kowalsky, Kronik, Snaith, Pucci, \latin{et~al.}
  others]{sendner2016optical}
Sendner,~M.; Nayak,~P.~K.; Egger,~D.~A.; Beck,~S.; M{\"u}ller,~C.; Epding,~B.;
  Kowalsky,~W.; Kronik,~L.; Snaith,~H.~J.; Pucci,~A. \latin{et~al.}  Optical
  phonons in methylammonium lead halide perovskites and implications for charge
  transport. \emph{Mater. Horiz.} \textbf{2016}, \emph{3}, 613--620\relax
\mciteBstWouldAddEndPuncttrue
\mciteSetBstMidEndSepPunct{\mcitedefaultmidpunct}
{\mcitedefaultendpunct}{\mcitedefaultseppunct}\relax
\EndOfBibitem
\bibitem[Cinquanta \latin{et~al.}(2019)Cinquanta, Meggiolaro, Motti, Gandini,
  Alcocer, Akkerman, Vozzi, Manna, De~Angelis, Petrozza, \latin{et~al.}
  others]{cinquanta2019ultrafast}
Cinquanta,~E.; Meggiolaro,~D.; Motti,~S.~G.; Gandini,~M.; Alcocer,~M.~J.;
  Akkerman,~Q.~A.; Vozzi,~C.; Manna,~L.; De~Angelis,~F.; Petrozza,~A.
  \latin{et~al.}  Ultrafast THz Probe of Photoinduced Polarons in Lead-Halide
  Perovskites. \emph{Phys. Rev. Lett.} \textbf{2019}, \emph{122}, 166601\relax
\mciteBstWouldAddEndPuncttrue
\mciteSetBstMidEndSepPunct{\mcitedefaultmidpunct}
{\mcitedefaultendpunct}{\mcitedefaultseppunct}\relax
\EndOfBibitem
\end{mcitethebibliography}

\begin{tocentry}
\includegraphics{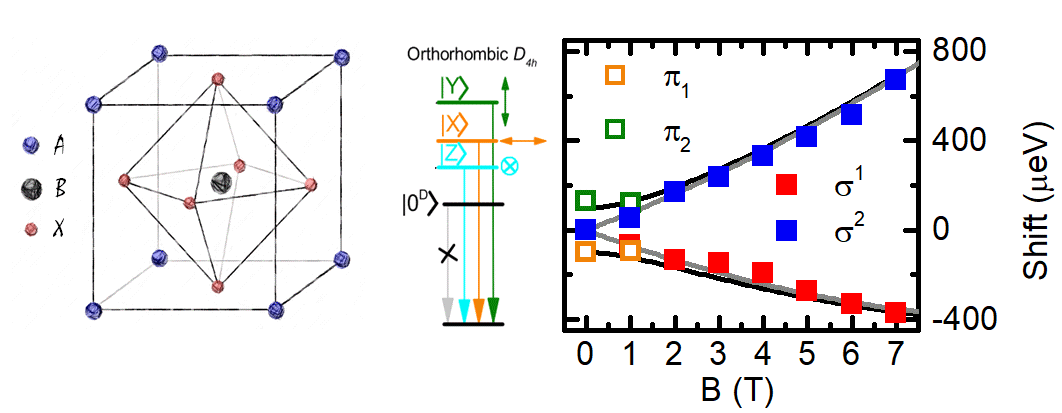}
\end{tocentry}

\end{document}